# Comparison of PM-HIP to Forged SA508 Pressure Vessel Steel Under High-Dose Neutron Irradiation

Wen Jiang [a+], Yangyang Zhao [a+], Yu Lu [b,c], Yaqiao Wu [b,c], David Frazer [d],

Donna P. Guillen [d], David W. Gandy [e], Janelle P. Wharry [a*]

[a] *School of Materials Engineering, Purdue University, West Lafayette, IN, USA*

[b] *Micron School of Materials Science & Engineering, Boise State University, Boise, ID, USA*

[c] *Center for Advanced Energy Studies, Idaho Falls, ID, USA*

[d] *Idaho National Laboratory, Idaho Falls, ID, USA*

[e] *Electric Power Research Institute, Charlotte, NC, USA*

+ These authors have equivalent contributions as first author

**\* Corresponding Author**

**Present/Permanent Address:**

Janelle Wharry,

Associate Professor, Purdue

jwharry@purdue.edu

701 W Stadium Boulevard

West Lafayette, IN, USA 47906

765-494-0782

**Abstract**

Powder metallurgy with hot isostatic pressing (PM-HIP) is an advanced manufacturing process that is envisioned to replace forging for heavy nuclear components, including the reactor pressure vessel (RPV). But PM-HIP products must at least demonstrate comparable irradiation tolerance than forgings in order to be qualified for nuclear applications. The objective of this study is to directly compare PM-HIP to forged SA508 Grade 3 Class 1 low-alloy RPV steel at two neutron irradiation conditions: ~0.5-1.0 displacements per atom (dpa) at ~270ºC and ~370ºC. PM-HIP SA508 experiences greater irradiation hardening and embrittlement (total elongation) than forged SA508. However, uniform elongation and approximate toughness are comparable across all irradiated materials, suggesting irradiated PM-HIP SA508 exhibits superior ductility at maximum load-bearing capacity. The irradiation hardening mechanism is linked to composition rather than fabrication method. Since PM-HIP SA508 has higher Mn and Ni concentration, it is more susceptible to irradiation-induced nucleation of Mn-Ni-Si-P (MNSP) nanoprecipitates and dislocation loops, which both contribute to hardening. Conversely, the forged material nucleates fewer MNSPs, causing dislocation loops to control irradiation hardening. These results show promise for the irradiation performance of PM-HIP SA508 and can motivate future nuclear code qualification of PM-HIP fabrication for RPVs.

# 1. Introduction

Nuclear reactor pressure vessels (RPV) are Class 1 safety structures preventing radioactive materials from entering the environment, and are thus of paramount importance to the safe operation and sustainability of the nuclear energy enterprise [1]–[3]. The RPV is irreplaceable, underscoring the importance of maintaining its structural integrity throughout its service life [1]–[9]. In commercial light water reactors (LWR), RPVs will typically experience extreme environments combining irradiation doses up to ~0.2-0.5 displacements per atom (dpa) by end-of-life, high temperatures ~280-320ºC, high pressures, mechanical stresses, and corrosive water [10]. These environments are even harsher in advanced Generation IV nuclear reactors systems [11]–[13]. During design-basis accident scenarios, the RPV must maintain sufficient strength to withstand internal pressure and temperature excursions, while maintaining high fracture and impact toughness during these events

Despite the paramount importance of the RPV to reactor safety, manufacturing of RPVs has been faced with numerous challenges. The first generation of RPVs were fabricated in sections, then joined using longitudinal and circumferential submerged-arc welds. But surveillance specimens revealed greater irradiation hardening and embrittlement along these weld seams [14]. To reduce the susceptibility of RPVs to catastrophic through-wall cracking [15]–[19], later RPVs were forged in a single piece, without welds. But globally, only a few facilities have the capability to forge an entire RPV in one piece, resulting in long lead times and exorbitant costs, making RPV fabrication the rate-limiting-step for new nuclear construction. If new nuclear power facilities are to be built to address global demands for low-carbon electricity, alternatives to forging for RPV manufacture must be identified.

Powder metallurgy with hot isostatic pressing (PM-HIP) is an advanced manufacturing technology that densifies metallic powders using a combination of high temperatures (~$0.7T_m$) and pressures (>100 MPa) [20]. PM-HIP is attractive for nuclear RPVs because throughput can be as high as that of metal injection molding, while components >10 ton can be produced (~5 orders of magnitude greater than the largest parts that can be fabricated by additive manufacturing). Additionally, PM-HIP components are fabricated near-net shape, reducing the need for post-processing, machining, and welding [21]–[23]. PM-HIP also eliminates casting and forging defects [24], [25] such as pinholes, blowholes, cold shuts, hot cracking, and surface cracking, rendering the components easier to inspect for quality. At the microstructural level, PM-HIP grains are equiaxed and refined [26], [27], with limited intergranular segregation [28]–[32], owing to the sub-melting temperature of the HIP process. This microstructural uniformity tends to give PM-HIP components superior as-fabricated mechanical properties compared to castings and forgings [33]–[37].

PM-HIP is qualified alongside casting and forging, as an acceptable fabrication route for ferritic and austenitic steel as well as Ni-based alloys for non-nuclear applications in the ASME

Boiler and Pressure Vessel Code (BPVC) Section II. The PM-HIP product of austenitic stainless steel 316L has recently become qualified for non-irradiation facing components in nuclear reactors through ASME BPVC Section III. The nuclear industry has growing interest in qualifying additional PM-HIP alloys for nuclear service, including for irradiation-facing environments [24], [25]. But nuclear code qualification requires demonstration of acceptable irradiation performance of PM-HIP components comparable to fabrication methods already qualified (i.e., casting or forging) [25].

Thus far, limited studies have performed side-by-side comparisons of the irradiation effects of PM-HIP and cast or forged nuclear structural alloys. Amongst the most systematic studies are those from Clement, et al. [26], [27], who compared the irradiation performance of PM-HIP and forged Alloy 625, a Ni-base alloy proposed for high temperature nuclear components. Clement's first study used ion irradiation to doses of 50 and 100 displacements per atom (dpa) to demonstrate that the initial dislocation density was the major controlling factor governing differences in the irradiated microstructure evolution between PM-HIP and forged variants. Later, Clement compared the same materials following 0.5-1 dpa neutron irradiation and found the PM-HIP Alloy 625 retained superior irradiated mechanical properties, including tensile strength and ductility [26]. The same neutron irradiation campaign [38] also included specimens of 316L austenitic stainless steel and Grade 91 ferritic steel. After neutron irradiation of almost 4 dpa at ~380ºC, the PM-HIP 316L exhibits lower irradiation hardening, greater strain hardening capacity, and greater ductility than its cast counterpart [39]. In the Grade 91 steel, Ni-Mn-Si-rich nanoprecipitates nucleate during ~1 dpa, 390ºC neutron irradiation, but these nanoprecipitates are smaller and less dense in the PM-HIP than in the cast variant [39]. From these limited data sets in high-alloyed materials, PM-HIP shows promise for favorable irradiation performance in comparison to its cast or forged counterparts.

The irradiation performance of RPV steels, typically low-alloy steels, has been the subject of tremendous research for more than a half-century [1], [3]. Phenomena such as irradiation hardening [40] and embrittlement [9], [41], [42] are well understood, and linked to irradiation-induced dislocation loops and nanoprecipitates. But to the authors' knowledge, there has only ever been one report of irradiation effects in PM-HIP RPV steels, from Carter et al. [43]. They conducted neutron irradiation of PM-HIP SA508 to 0.1 dpa at ~155 °C, then used nanoindentation to identify greater hardening in the ferrite phases than in the bainite phases, attributed to the higher tendency for solute nanoprecipitation in ferrite [43]. Nevertheless, these irradiation effects in PM-HIP SA508 have not yet been directly benchmarked against a cast or forged materials.

The objective of the current study is to compare side-by-side the neutron irradiation performance of PM-HIP to forged SA508 Grade 3 RPV steel. SA508 is a Ni and Mo modified low-alloy steel developed for its formability, but early grades exhibit limited toughness and suffer from reheat cracking [44]. SA508 Grade 3 has lower C, Cr, and Mo content to limit reheat cracking and Mn additions for improved strength [45]; it is standard RPV steel in the United States and China. In this study, PM-HIP and forged variants of SA508 Grade 3 are neutron irradiated to nearly identical conditions over the dose range ~0.5-1 dpa and temperature range ~270-400 ºC. Uniaxial tensile testing and fractography provide insight into the mechanical implications of irradiation. The irradiated microstructure is characterized across a range of length scales, spanning from the grain/phase scale through the nanoprecipitate or solute nanocluster length scale. The structure-property relationships are established for both the PM-HIP and forged variants, and microstructural mechanisms underlying the relative irradiation susceptibilities of both materials are discussed. The paper will conclude with a discussion on the readiness of PM-HIP RPV steels for nuclear qualification.

## 2. Experimental methods

### 2.1 Materials and irradiation

Ingots of PM-HIP and forged SA508, Grade 3, Class 1 steels were provided by the Electric Power Research Institute (EPRI). To fabricate the PM-HIP ingot, gas atomized SA508 powders were mixed and filled into a canister that was outgassed, welded closed, then underwent HIP at a 103 MPa pressure at 1121 °C for 4 hours. Subsequently, the PM-HIP ingot was solution annealed (1121 °C for 2 hours followed by water quenching), normalized (899 °C for 10 hours followed by water quenching), and subsequently tempered (649 °C for 10 hours followed by air cooling). The forged ingot underwent the identical solution anneal, normalization, and tempering treatment. The chemical compositions of the PM-HIP and forged samples were analyzed using inductively coupled plasma atomic emission spectroscopy (ICP-AES), Table 1. Both materials are consistent with the ASTM specification for SA508 Grade 3, except for having lower Mo composition.

*Table 1 Chemical compositions (in wt%) of the investigated SA508 specimens, compared to ASTM specification for SA508 Grade 3.*

| Alloy | C | Si | Mn | Ni | Cr | Mo | V | P | S | Fe |
|---|---|---|---|---|---|---|---|---|---|---|
| PM-HIP | 0.01 | 0.21 | 1.39 | 0.79 | 0.18 | 0.37 | - | 0.002 | 0.005 | Bal. |
| Forged | 0.02 | 0.31 | 1.46 | 0.50 | 0.21 | 0.26 | 0.01 | 0.003 | 0.007 | Bal. |
| ASTM Specification | <0.25 | 0.15-0.40 | 1.20-1.50 | 0.40-1.00 | <0.25 | 0.45-0.60 | <0.05 | <0.025 | <0.025 | Bal. |

The PM-HIP and forged ingots were sectioned into two different specimen geometries for neutron irradiation: round tensile bars and circular discs. Round tensile specimens conformed to ASTM E8 standards, with a nominal gauge length of 31.75 mm and diameter of 6.35 mm. The tensile bars were machined at Idaho National Laboratory (INL) using computer numerical control machining to a surface roughness of 3.2 µm. Disc-shaped specimens had a diameter of 3 mm and thickness of 0.15 mm. The discs were prepared at INL using wire electrical discharge machining with subsequent hand polishing to a mirror finish. Comprehensive specimen drawings are provided in ref. [46].

The tensile bars and disc specimens were loaded into drop-in capsules for irradiation in the Advanced Test Reactor (ATR) at INL. Irradiation was performed during ATR cycle 164A, with target irradiation temperatures of 300 °C and 400 °C, and a target dose of 1 displacement per atom (dpa). A general-purpose Monte Carlo N-Particle (MCNP) transport code, MCNP5 release 1.40, was used to calculate the actual as-run irradiation doses accumulated on the specimens [47]. The total neutron flux was $8.1 \times 10^{14}$ n/cm$^2$·s, with fast (>1 MeV) flux component of $1.6 \times 10^{14}$ n/cm$^2$·s. The effect of γ flux was included as γ-heating in the target temperature estimation but was not considered in the dose estimate since photons generally create negligible displacement damage in metallic alloys. Finite element analysis was performed using ABAQUS to determine the actual as-run temperatures experienced by the specimens.

Four irradiated disc specimens and six irradiated tensile bars were selected for microstructural and mechanical characterization, respectively; unirradiated tensile and disc specimens were also studied as controls. The actual doses and temperatures for each specimen are given in Table 2. Actual doses ranged 0.53-1.00 dpa, with an average dose of 0.83±0.17 dpa, corresponding to a dose rate range of $1.0$-$1.8 \times 10^{-7}$ dpa/s. The target 300 ºC specimens experienced an actual temperature range of 265-286 ºC, with an average temperature of 274±9 ºC. The target 400 ºC specimens experienced a temperature range of 343-388 ºC, with an average temperature of 371±16 ºC. Henceforth, the actual as-run doses and temperatures will be used to describe specimens. Details of the irradiation experiments, including the dose and temperature calculations, and the meaning of the specimen ID numbers given in Table 2, are comprehensively explained in [46].

*Table 2* Summary of neutron irradiation conditions for the PM-HIP and forged SA508 specimens (specimens marked with * indicate pairs of duplicate specimens intended for statistical confidence).

| Alloy | Type | Specimen ID (see [46]) | Target dose (dpa) | Target temp. (ºC) | Actual dose (dpa) | Actual average temp. (°C) | Actual maximum temp. (°C) |
|---|---|---|---|---|---|---|---|
| PM-HIP | Tensile | 101 | 1 | 300 | 0.54 | 266 | 273 |
| PM-HIP | Disc | 104 | 1 | 300 | 0.69 | 286 | 286 |
| PM-HIP | Tensile* | 401 | 1 | 400 | 0.97 | 343 | 360 |
| PM-HIP | Tensile* | 402 | 1 | 400 | 1.00 | 365 | 375 |
| PM-HIP | Disc | 431 | 1 | 400 | 0.97 | 388 | 389 |
| Forged | Tensile | 206 | 1 | 300 | 0.83 | 270 | 281 |
| Forged | Disc | 110 | 1 | 300 | 0.69 | 286 | 286 |
| Forged | Tensile* | 501 | 1 | 400 | 0.96 | 362 | 379 |
| Forged | Tensile* | 502 | 1 | 400 | 0.98 | 385 | 395 |
| Forged | Disc | 437 | 1 | 400 | 0.95 | 384 | 384 |

## *2.2 Mechanical testing*

Quasi-static uniaxial tensile testing of the unirradiated and irradiated specimens were carried out in accordance with ASTM standard E8. Tensile testing was conducted on the Remote Operated Instron 5869 screw-driven load frame at the Hot Fuel Examination Facility (HFEF) Main Cell Window 13M at the Materials and Fuels Complex (MFC) at INL. For the round tensile bars herein, threaded buttons are used on both ends of the tensile bar, onto which the instrument can grip. Tensile tests were conducted at room temperature in an argon (Ar) environment with a crosshead speed of 0.279 mm min$^{-1}$, corresponding to a strain rate of $1.5 \times 10^{-4}$ s$^{-1}$. Complete details of the tensile testing are provided in [48].

Before conducting the tensile tests reported herein, two redundant irradiated specimens of the PM-HIP material were placed in the Instron load frame. Both of those specimens broke at the threads due to their irradiation embrittlement, rendering them untestable and essentially wasting these specimens. To preserve the remaining irradiated specimens without risking them fracturing at the threads, the threaded buttons were placed lower on the tensile bars than usual and the extensometer was placed on the grips. This allowed for accurate measurement of yield strength and ultimate tensile strength, but compromised the Young's modulus measurements. Additionally, ductility measurements needed to be adjusted for compliance of the grip.

Irradiation hardening ($\Delta\sigma_y$) was determined from the resultant stress-strain curves as the increment in yield strength induced by irradiation, that is, $\Delta\sigma_y = \sigma_{y,irr} - \sigma_{y,unirr}$, where $\sigma_{y,irr}$ and $\sigma_{y,unirr}$ are the yield strength before and after irradiation, respectively. Following tensile testing, fractography was conducted using a Tescan Lyra3 scanning electron microscope (SEM) at HFEF at INL.

## *2.3 Scanning and transmission electron microscopy*

Unirradiated control specimens were prepared for microscopy by mechanical grinding with SiC paper, polishing through 1 μm diamond paste, electropolishing, then etching. Electropolishing was conducted using a solution of $HClO_4:C_2H_6O$ at a 1:9 volume ratio, at 20 V and a temperature of -25 °C using a Buehler ElectroMet-4 polisher. Etching was conducted by immersion in 4 vol.% Nital. Preparation of the irradiated disc samples for microscopy involved electrochemical polishing in a solution of HClO$_4$:C$_2$H$_6$O at a 1:7 volume ratio, at 18 V and a temperature of -17.5 °C using a Struers TenuPol-5 electropolishing system at the MFC of INL. The disc specimens were then sent to the Microscopy and Characterization Suite (MaCS) at the Center for Advanced Energy Studies (CAES) for SEM and TEM characterization.

Grain and phase structural characterization was conducted using SEM in bacskscatter electron (BSE) mode. Ferrite and bainite phases were distinguished based on their contrast in BSE mode, in which ferrite phases present as dark contrast while bainite phases are decorated with light-contrasting proeutectoid cementite [49]. Grain and phase sizes were determined using ASTM E112-13 on a minimum of 10 images from each material condition. For the unirradiated control specimens, this SEM characterization was conducted using a ThermoFisher Helios G4 UX dual-beam SEM/focused ion beam (FIB) at Purdue University. For the irradiated specimens, this SEM characterization was conducted using a ThermoFisher (formerly FEI) Quanta 3D FEG dual-beam SEM/FIB at MaCS, CAES.

Lamellae for TEM analysis were extracted from both the unirradiated and irradiated specimens using the FIB lift-out method following typical protocols [50], [51]. Before FIB cutting, a platinum layer was deposited on the sample surface to prevent possible surface damage caused by Ga$^+$ implantation and sputter erosion during ion milling. To minimize FIB-induced damage, low energy cleaning using a 2 keV Ga$^+$ beam was applied to TEM lamellae as the final preparation step. Unirradiated FIB work was conducted on the Helios G4 UX SEM/FIB at Purdue, while the irradiated FIB work was conducted on the Quanta 3D FEG SEM/FIB at MaCS, CAES.

TEM characterization focused on precipitates and dislocation loops. Precipitate characterization was conducted using high resolution scanning transmission electron microscopy (STEM), with fast Fourier transformation (FFT) to discern crystallinity. Note the precipitates were too small to identify their crystal structures via selected area electron diffraction (SAED). Dislocation loops were imaged using the bright field (BF) STEM technique [52]–[55] due to its relaxed diffraction conditions caused by the convergent beam under BF-STEM, enabling simultaneous imaging of all loop orientations, thus simplifying the quantification of the total loop number density. This technique allows for rapid imaging with strong contrast, as compared to conventional methods (e.g., two-beam and weak-beam dark field imaging [56], [57]), as well as high statistical accuracy [58]–[60]. In this work, BF-STEM micrographs were consistently collected from the [011] zone axis for quantitative analysis of dislocation loops. To calculate volumetric loop number densities, the average thickness of each TEM lamella was measured by electron energy loss spectroscopy in energy-filtered TEM mode [61]. Voids were sought using through-focus BF-TEM imaging, meaning that voids (if present) would be identifiable as features exhibiting opposite contrast in over-focused conditions than in under-focused conditions. However, no voids could be identified. The unirradiated specimen TEM work was done on a ThermoFisher (formerly FEI) Talos 200X transmission electron microscope (TEM) operated at 200 kV equipped with a high-angle annular dark field (HAADF) detector and a Super-X energy dispersive x-ray spectroscopy (EDS) system at Purdue University. The irradiated TEM work was conducted using a ThermoFisher (formerly FEI) Tecnai TF30-FEG STwin TEM operating at 300 kV equipped with a HAADF detector for use in STEM mode at MaCS, CAES. Unirradiated specimens were also imaged by BF-STEM on the ThermoFisher Tecnai TEM at MaCS, CAES, to establish a baseline to ensure that features counted as irradiation-induced dislocation loops are not merely FIB artifacts.

*2.4 Atom probe tomography*

The electropolished irradiated disc specimens and unirradiated control specimens used for TEM characterization were also used for atom probe tomography (APT). Similar FIB lift-out procedures were followed to prepare APT needles on the ThermoFisher Quanta 3D FIB/SEM at MaCS. After lift-out of the lamella, it was sliced into ~6 needles, each of which was welded onto a Si post on an APT coupon. Top-down annular FIB milling was used to shape each needle to a final diameter of ~50 nm at the tip. To remove FIB damage introduced during annular milling, low energy annular cleaning using a 2 keV Ga$^+$ beam was applied as the final step.

APT needles were run on a Cameca LEAP 4000X HR local electrode atom probe operated in pulsed-laser mode at a specimen temperature of 50 K, with a pulse repetition rate of 200 kHz and a focused laser beam energy of 60 pJ. Detection rate was between 0.5% and 1% of all ions per field evaporation pulse. Datasets containing 10-20 million ions were acquired from a minimum of two needles for each irradiation condition. APT data reconstruction and analysis were carried out using the Cameca APSuite commercial software. Cluster analysis was performed using the maximum separation method [62], [63] to estimate the size, volume fraction, and number density of nanoprecipitates. In the cluster analysis, parameters of $d_{max}$ = 0.57-0.87 nm and $N_{min}$ = 12-27 were used, where $d_{max}$ is the maximum separation distance between solute atoms for cluster identification and $N_{min}$ is the minimum number of solute atoms in a cluster. The cluster size ($r$) and number density ($n_v$) were determined following standard procedures established in ref. [64], [65]; these methods are described comprehensively in Item S1 in the Supplementary Information

## 3. Results

*3.1 Tensile properties & fractography*

The room-temperature engineering stress–strain curves of the unirradiated and irradiated PM-HIP and forged SA508 pressure vessel steels are presented in Fig. 1(a). Tensile properties including yield strength (YS), ultimate tensile strength (UTS), uniform elongation (UE), total elongation (TE), toughness (T), and irradiation hardening ($\Delta\sigma_y$) determined from these stress-strain curves are summarized in Table 3 and compared visually in Fig. 1(b-d). The mechanical behavior of the PM-HIP material is striking even before irradiation. The unirradiated PM-HIP specimen exhibits ~120 MPa higher YS and ~180 MPa higher UTS (Fig. 1(b)) without significant loss of UE and TE (Fig. 1(c)) as compared to the reference forged material. Mechanical behaviors between PM-HIP and forged materials deviate even more after irradiation.

Exposure of both the PM-HIP and forged materials to irradiation consistently results in yield and tensile strength increases (Fig. 1(b)), leading to hardening (Fig. 1(c)), with a reduction in ductility (Fig. 1(d)). The PM-HIP specimens show significantly higher strength increases than those of their conventional forged counterparts. Specifically, PM-HIP YS and UTS increase by ~450-600 MPa, while forged YS and UTS increase by only ~190-280 MPa, across all irradiation conditions studied. Collectively, these results suggest more severe irradiation hardening in the PM-HIP than in the forged material.

The effects of irradiation on ductility and toughness, however, do not necessarily follow irradiation hardening trends. That is, TE of the irradiated PM-HIP specimens (6.3-9.6%) is almost half the TE of the irradiated forged specimens (12.1-16.6%). However, UE after irradiation is relatively consistent across both fabrication methods, ranging 4.2-5.2% in the PM-HIP specimens and 4.7-7.2% in the forged specimens. These results point to two key trends. First, under irradiation, TE to fracture is severely compromised in the PM-HIP material compared to the forging. But second – and more critical to the establishment of reactor operational safety margins – the irradiated PM-HIP and forging exhibit comparable ductility through their respective maximum load-bearing capacities. In addition, the approximate toughness (determined by integrating area under the stress-strain curves) is relatively comparable between the PM-HIP and forged specimens across all irradiation conditions, ranging ~0.74-1.17 × $10^8$ J/m$^{-3}$ for PM-HIP compared to ~0.82-1.12 × $10^8$ J/m$^{-3}$ for the forged specimens. Although PM-HIP SA508 may be more susceptible to irradiation hardening and reduction in TE than the forged material, the PM-HIP exhibits comparable resistance to fracture and embrittlement at maximum load.

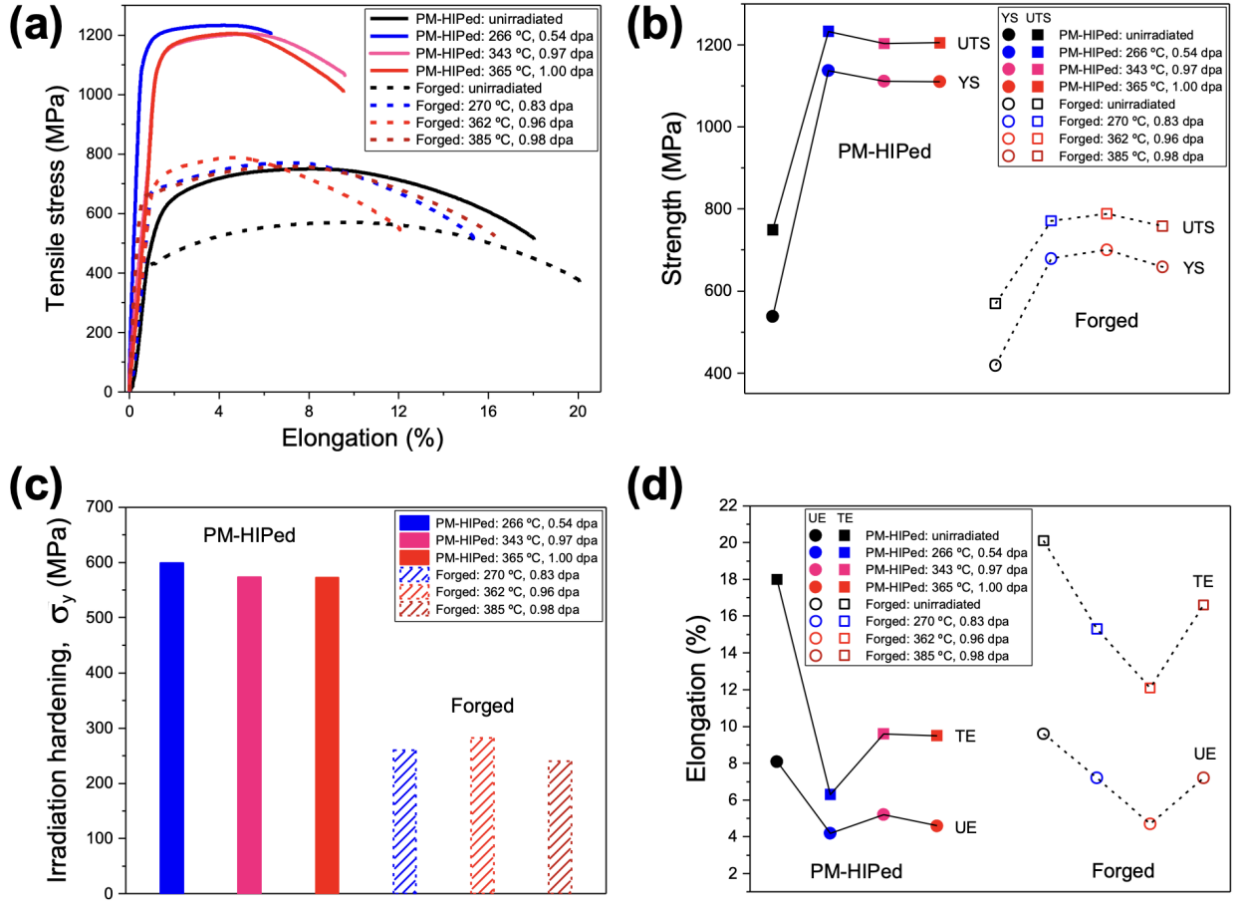

*Fig. 1* *(a) Room temperature tensile engineering stress-strain curves for the PM-HIP and forged SA508 samples exposed to different irradiation conditions. Overview of the quantitative tensile mechanical properties, specifically (b) yield strength (YS) and ultimate tensile strength (UTS), (c) irradiation hardening ($\Delta\sigma_y$), and (d) uniform elongation (UE) and total elongation (TE).*

Mechanical property trends as a function of irradiation dose or temperature are difficult to extract since limited data are available. In the PM-HIP material, YS, UTS, and irradiation hardening, irradiated YS, and irradiated UTS, all tend to increase with decreasing irradiation temperature. This trend is evident despite the dose at the lowest temperature (266 °C, 0.54 dpa) being almost half the dose at the two higher temperatures (343 and 365 °C, ~1 dpa). Meanwhile, in the forged material, YS, UTS, and irradiation hardening appear to peak at the middle temperature (362 °C, 0.96 dpa), although this may partly be attributed to the dose being marginally lower at the lowest temperature (270 °C, 0.83 dpa). Elongation and hardening are inversely related in both the PM-HIP and forged materials; that is, specimens with high irradiation hardening tend to have lower elongation, and vice versa. Hence, TE and UE trends as a function of irradiation temperature are mirror opposites of those described for hardening. In sum, these trends suggest the possibility that the irradiation hardening and embrittlement regime may shift to lower temperatures for PM-HIP compared to forged materials, although one must also consider the context that standard deviations on as-run temperatures are ±39 °C [46].

*Table 3 Room temperature tensile properties of unirradiated and irradiated PM-HIP and forged SA508.*

| Alloy | Irradiation Conditions | YS (MPa) | UTS (MPa) | UE (%) | TE (%) | T ($10^8$ J/m$^{-3}$) | $\Delta\sigma_y$ (MPa) |
|---|---|---|---|---|---|---|---|
| PM-HIP | unirradiated | 538 | 750 | 8.1 | 18.0 | 1.17 | - |
| PM-HIP | 266 °C, 0.54 dpa | 1137 | 1233 | 4.2 | 6.3 | 0.74 | 599 |
| PM-HIP | 343 °C, 0.97 dpa | 1111 | 1203 | 5.2 | 9.6 | 1.00 | 573 |
| PM-HIP | 365 °C, 1.00 dpa | 1111 | 1205 | 4.6 | 9.5 | 0.98 | 573 |
| Forged | unirradiated | 419 | 570 | 9.6 | 20.1 | 1.00 | - |
| Forged | 270 °C, 0.83 dpa | 679 | 771 | 7.2 | 15.3 | 1.02 | 260 |
| Forged | 362 °C, 0.96 dpa | 701 | 788 | 4.7 | 12.1 | 0.83 | 282 |
| Forged | 385 °C, 0.98 dpa | 659 | 758 | 7.2 | 16.6 | 1.12 | 240 |

Fig. 2 shows the fracture surfaces of PM-HIP and forged SA508 following 343 °C, 0.97 dpa (specimen ID 401) and 385 °C, 0.98 dpa (specimen ID 502) irradiation. The fracture surface of the PM-HIP material exhibits a relatively flat surface characteristic of brittle fracture; higher magnification imaging shows the fracture surface is dimple-free. Meanwhile, the forged material exhibits a combination of a relatively flat region in the center of the tensile bar with cup-cone like features around the periphery of the bar, suggesting a combination of brittle and ductile fracture modes. Higher magnification imaging reveals some dimpling. These fracture surface characteristics corroborate the tensile testing results in which this PM-HIP tensile bar exhibits 573 MPa irradiation hardening and 9.6% TE compared to this forged tensile bar which exhibits only 240 MPa hardening and 16.6% TE.

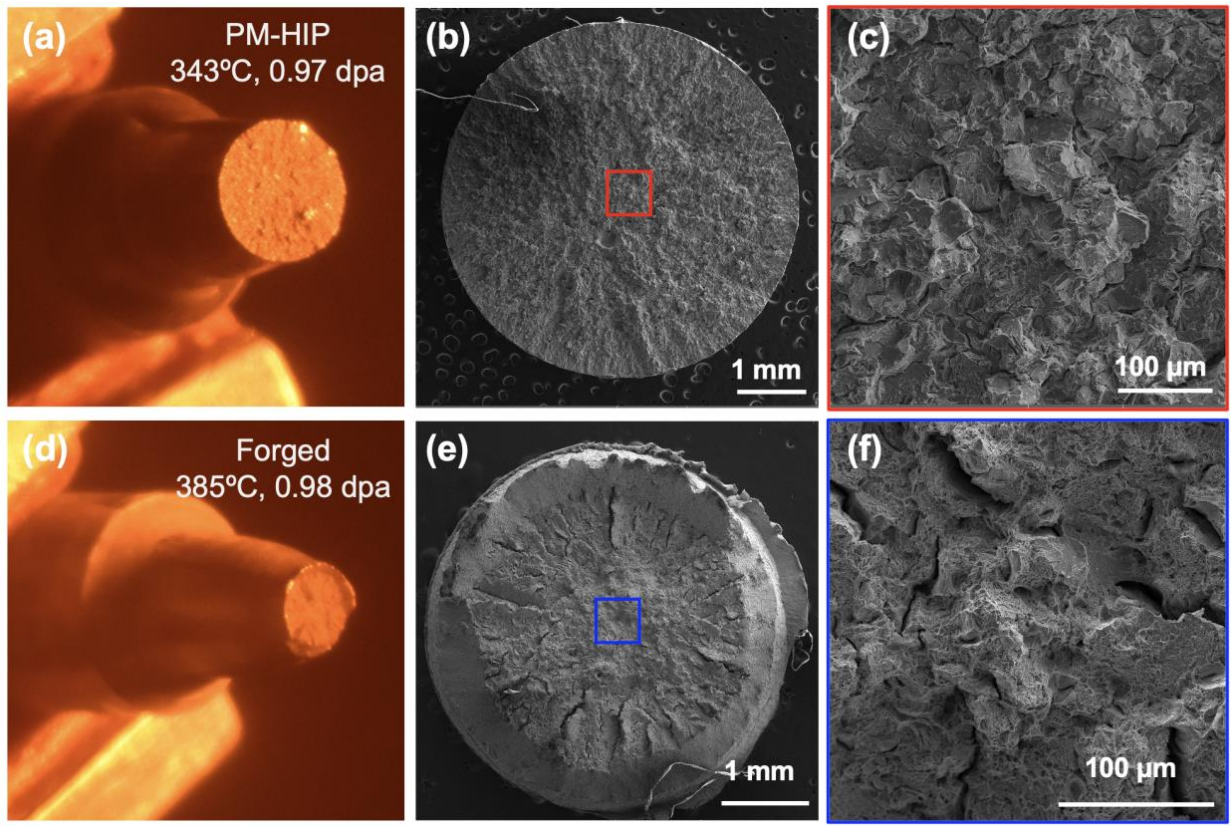

***Fig. 2*** *Fracture surface images of (a-c) PM-HIP SA508, following 343 ºC, 0.97 dpa irradiation (specimen ID 401) and (d-e) forged SA508, following 385 ºC, 0.98 dpa irradiation (specimen ID 502). For each fabrication method, the series of images are shown in order of increasing magnification, from the hot cell camera, SEM, and higher magnification SEM.*

To the authors' knowledge, only one previous publication has characterized the irradiation hardening of PM-HIP RPV steel. Carter et al. [43] measures an average $\sigma_y$ increase of ~56% (i.e., 1.94 GPa irradiation hardening measured by nanoindentation) for PM-HIP SA508 steel after 0.1 dpa neutron irradiation at ~155 °C. By contrast, the irradiation hardening of a conventionally fabricated RPV steel irradiated to the identical dose of 0.1 dpa at 290-295 ºC experiences only 29% increase in $\sigma_y$ (~140 MPa measured by uniaxial tensile testing) [66]. Recall the results of the present study show ~106% increase in $\sigma_y$ (~570-600 MPa hardening) in the PM-HIP alloy, compared to ~60% increase in $\sigma_y$ (~240-280 MPa hardening) in the forged alloy. Although the magnitudes of hardening herein exceed those in the aforementioned references – by virtue of their ~10x greater irradiation dose – a consistent trend emerges for PM-HIP RPV steels to exhibit more irradiation hardening than conventionally fabricated RPV steels under similar irradiation conditions.

Limited studies have measured irradiation hardening in RPV steels at doses in excess of LWR lifetime doses, i.e., ~0.15 dpa. One such study from Byun and Farrell, uses uniaxial tensile testing to measure irradiation hardening of ~450-550 MPa for A533B RPV steel neutron irradiated to ~1 dpa at ≤200 ºC [67]. These values fall into reasonable agreement with hardening magnitudes at similar doses in the present study, which is surprising since irradiation hardening is generally more extreme at ≤200 ºC than at the temperatures studied here. However, the similar hardening values may be reconciled by the higher dose rate in the current study, which limits the extent of point defect recombination at sinks, leaving more defects (and defect clusters) in the matrix to cause hardening in the present study. Additionally, Byun and Farrell suggest that irradiation hardening saturates by a dose of only 0.03-0.05 dpa [67]. These saturation doses are consistent with those predicted by Odette, Morgan, and coworkers using classical empirical models as well as modern machine learning models of irradiation hardening [1], [68]–[70]. Despite these saturation predictions also being at lower temperatures and dose rates than in the present study, saturation of hardening is likely by the doses herein.

*3.2 Phases and precipitation*

The unirradiated PM-HIP and forged SA508 both exhibit a dual-phase microstructure comprised of ferrite ($\alpha$) and bainite ($\alpha_\beta$) grains, which evolve during irradiation as shown in Fig. 3(a-c) for PM-HIP and Fig. 3(d-f) for forged. The PM-HIP material is comprised of 74% bainite, 26% ferrite, while the forged material is comprised of 49% bainite, 51% ferrite. In both materials, the bainite phase fraction tends to decrease with irradiation, while the ferrite phase fraction increases, as summarized in Table 4 and illustrated in Fig. S1 of Item S2 in the Supplementary Information. Initially, the PM-HIP material has bainite grain size of ~8.8 µm, and ferrite grain size of ~3 µm, whereas the initial forged material has statistically identical bainite and ferrite grain sizes ~10 µm. After irradiation, bainite grain size tends to decrease in both materials, Fig. 3(g), while ferrite grain size tends to increase, Fig. 3(h). But because phase size and phase fraction evolve simultaneously during irradiation, the overall grain size (i.e., weighted average of phase sizes) remains statistically unchanged throughout irradiation, Fig. 3(i). Specifically, the PM-HIP material overall grain size ranges from 4.68 to 6.08 µm, while the forged material overall grain size ranges from 10.1 to 11.2 µm. Since the ferrite phase exhibits greater irradiation hardening than the bainite phase [43], the remainder of microstructure characterization in this study focuses on the ferrite phase.

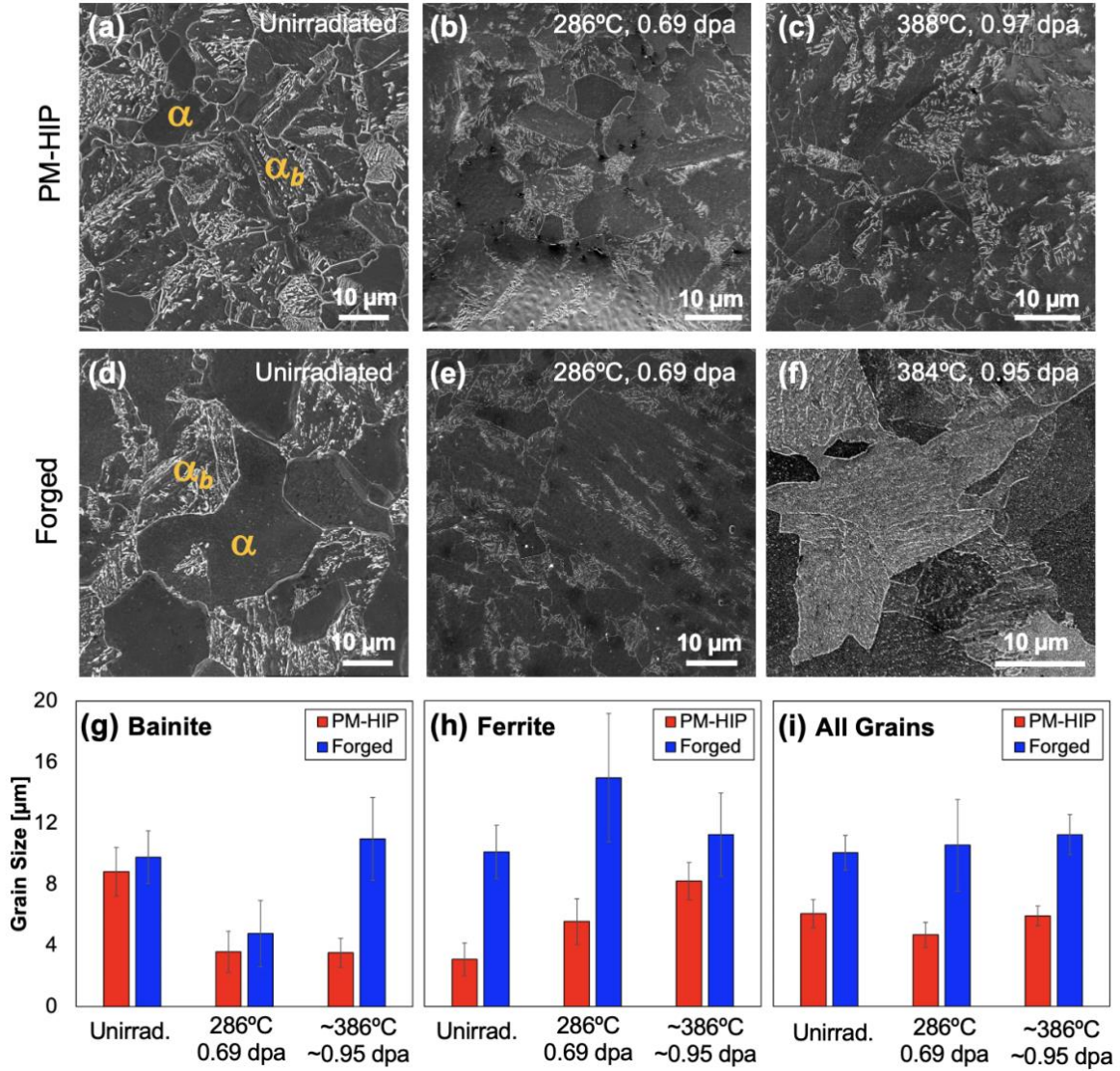

*Fig. 3* SEM images showing ferrite and bainite phase microstructure in unirradiated and irradiated (a-c) PM-HIP and (d-f) forged SA508, with quantitative evolution of sizes of (g) bainite grains, (h) ferrite grains, and (i) all grains across all material conditions.

Higher magnification TEM imaging and EDS line scans in the ferrite phases reveals distinct precipitation in the PM-HIP compared to forged specimens, Fig. 4. The needle-like precipitates in forged SA508 have average length 101 ± 6 nm, Fig. 4(d), and are believed to be $Mo_2C$ precipitates typical in bainitic ferrite [5], [71]. Fig. S2(a-c) of Item S2 in the Supplementary Information shows a ferrite-bainite interface with an EDS line scan over a selected precipitate. Meanwhile, the PM-HIP SA508 exhibits spheroidal Mn-Al-rich oxides of average diameter 49 ± 6 nm, Fig. 4(a) (micrographs and EDS line scan are provided in Fig. S2(a-c) of Item S2 in the Supplementary Information). Similar oxides have also been observed in PM-HIP SA508 by Carter et al, who suggests these oxides are induced by the HIP manufacturing process [43]. Retained oxide precipitates are not uncommon in PM-HIP metals [72]–[75], and often take the form of spheroidal precipitates [43]. The precipitates in Forged SA508 tend to be heterogeneously distributed, principally along grain boundaries, while the precipitates in PM-HIP SA508 are more uniformly dispersed throughout the grains. The average precipitate sizes in the forged and PM-HIP materials are statistically unchanged by both 300 °C and 400 °C irradiation, as reported in Table 4.

TEM characterization of the spherical oxide and needle-like precipitates reveals greater insight into their morphological and structural evolution under irradiation, Fig. 5. Prior to irradiation, both the PM-HIP and forged precipitates are crystalline, as confirmed by lattice fringes in the PM-HIP oxide as well as defined diffraction spots in the fast Fourier transformations (FFT) of both the oxide and needle-like precipitate. After irradiation to 0.95 dpa, 384 ºC, the needle-like precipitate in the forged material retains its diffraction spots indicative of crystallinity. However, after 0.97 dpa, 388 ºC irradiation, the oxide precipitate in the PM-HIP material exhibits amorphization.

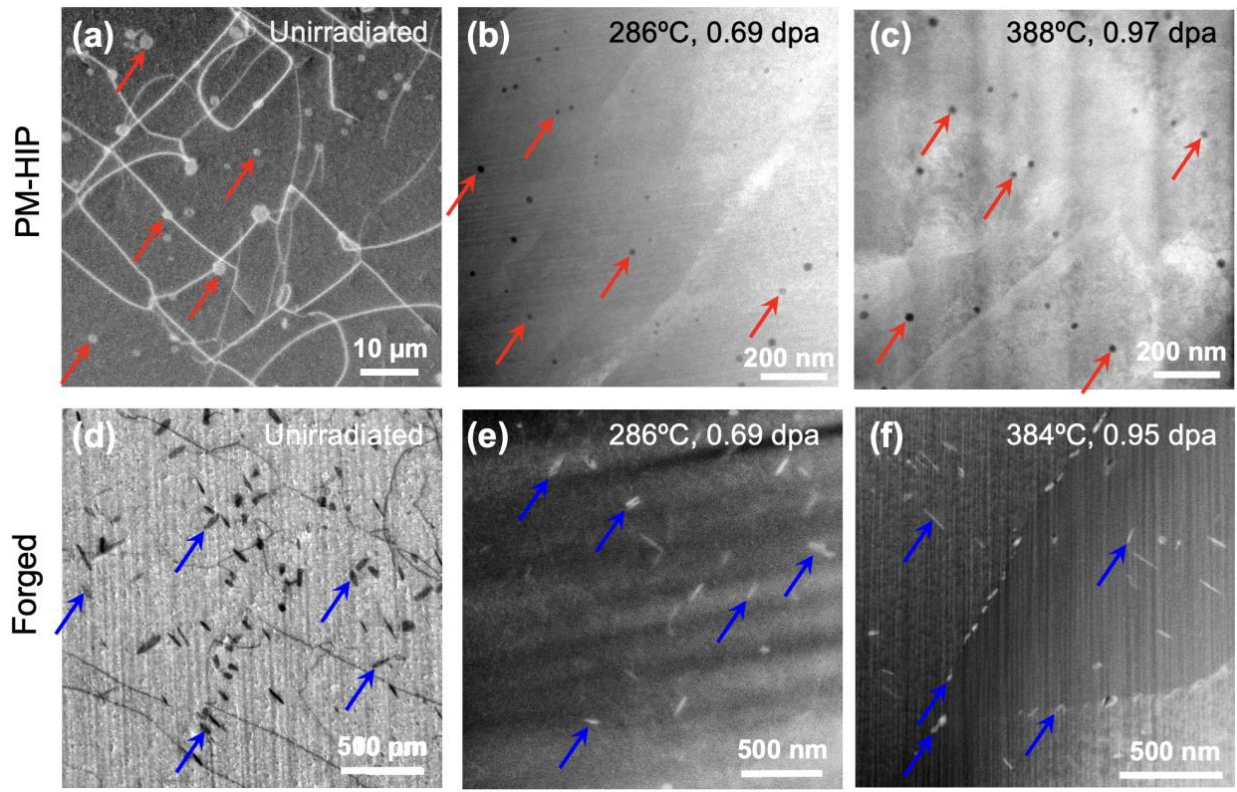

*Fig. 4* TEM micrographs of residual processing-induced precipitates in (a-c) PM-HIP and (d-f) forged SA508. (a,d) Unirradiated micrographs taken by STEM; (b,c,e,f) irradiated micrographs taken by HAADF. Example precipitates marked by arrows.

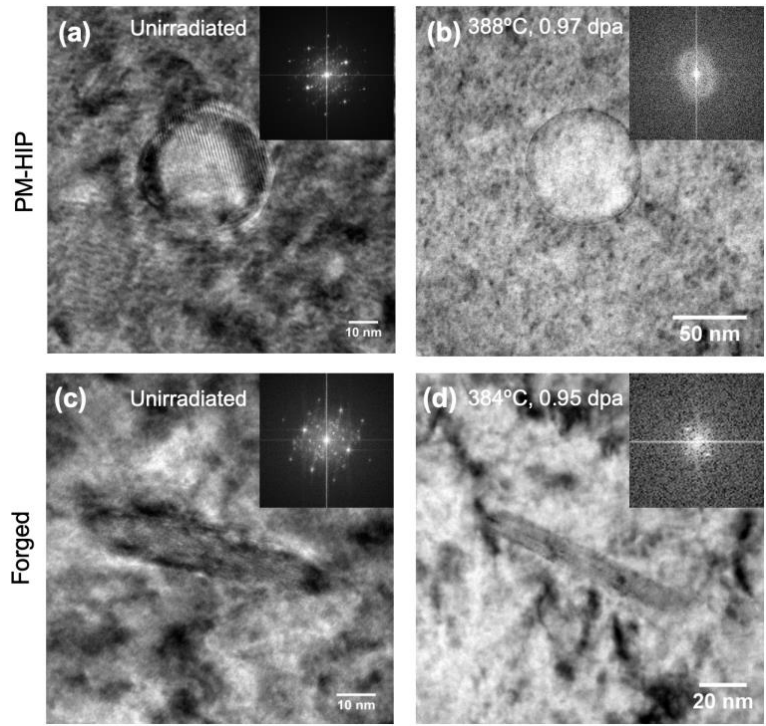

*Fig. 5 TEM micrographs of residual processing-induced precipitates in (a) unirradiated PM-HIP, (b) higher temperature irradiated PM-HIP, (c) unirradiated forged, and (d) higher temperature irradiated forged SA508. Insets show FFTs, indicating irradiation-induced amorphization of spherical precipitates in PM-HIP SA508, while needle-like precipitates in forged SA508 retain crystallinity during irradiation.*

*3.3 Dislocation loops*

Fig. 6 shows the evolution of dislocation loops in neutron irradiated PM-HIP and forged SA508 at all irradiation conditions studied. Unirradiated specimens are also shown under identical imaging conditions to demonstrate that features counted as irradiation-induced dislocation loops (arrowed in Fig. 6) are not FIB artifacts. For both alloys, loop sizes increase and number densities decrease with increasing irradiation temperature, Table 4. Quantitatively, the loop size for PM-HIP SA508 at 0.69 dpa, 286 °C is roughly half that of forged SA508 at the same irradiation conditions (3.90 nm compared to 6.39 nm). At ~0.95 dpa, ~385 °C, loop sizes in both alloys increases by a factor of about 1.5, which is likely attributed to both the higher temperature and dose. Meanwhile, the number density for PM-HIP SA508 at 0.69 dpa, 286 °C is ~8 times that of forged SA508 ($15.04 \times 10^{22}$ m$^{-3}$ compared to $1.81 \times 10^{22}$ m$^{-3}$) under the same conditions. Number densities in both alloys decrease by nearly half at ~0.95 dpa, ~385 °C.

These loop number densities are an order of magnitude (or more) higher than those reported in the vast majority of RPV steel literature, typically on the order of ~$10^{20}$-$10^{21}$ m$^{-3}$ at irradiation temperatures ~300-400°C (e.g., refs. [76]–[80]). However, much of the RPV steel literature considers LWR lifetime-relevant irradiation doses, ~0.01-0.15 dpa. Only a few studies have reported on microstructure evolution in RPV steels at doses ≳0.5 dpa. One such study from Jiang, et al. [81] reports 3.9 ± 1.2 nm diameter loops at a density of 3.7 × $10^{22}$ m$^{-3}$ in A508-3 RPV steel after 1.0 dpa, 240 keV proton irradiation at ~100 °C. Similarly, Fujii and Fukuya [21] and Watanabe, et al. [64] conduct 2.4-3 MeV heavy ion irradiation of A533B RPV steel to doses of 1-2 dpa at 290 °C and observe 2-2.5 nm diameter loops at number densities of 1-3 × $10^{22}$ m$^{-3}$. Finally, black dots (likely loop nuclei) are reported in A508-3 at a number density of 1.8 × $10^{22}$ m$^{-3}$ after room temperature 1.5 dpa Fe$^+$ ion irradiation [82]. Thus, the loop sizes and number densities reported herein are consistent with those observed at comparable irradiation doses.

Loop measurements reported here can also be considered in the context of other alloys irradiated in the same campaign. Clement et al. [26] studied PM-HIP and forged Ni-base Alloy 625 irradiated to ~1 dpa at 385 °C: conditions essentially identical to the higher-temperature specimens herein. Clement's Alloy 625 loop diameters range from 4.8 to 8.1 nm, which is comparable to those in SA508. However, loop number densities in Alloy 625 are roughly two orders of magnitudes smaller than in SA508 loops ($10^{20}$ m$^{-3}$ compared to $10^{22}$ m$^{-3}$) despite the alloys being identically irradiated. This is unsurprising, since higher-alloyed materials are known to exhibit greater resistance to irradiation damage than low-alloy materials or pure metals, due to the ability of impurity and minor elements to trap point defects and stabilize extended defects [83]–[85]. An extreme manifestation of this behavior is the recent observation of irradiation tolerance in high-entropy alloys, owing to the more sluggish diffusion kinetics and distorted energy landscape resulted from the compositional complexity [86]–[89].

Voids are not observed in either the forged or PM-HIP material, at both irradiation conditions studied. Although voids had been believed to play a role in irradiation hardening of RPV steels [90], they generally do not exist save for small clusters containing a few vacancies and are often associated with CRPs [66], [91]–[93], Thus, the absence of voids in the present work is not unusual.

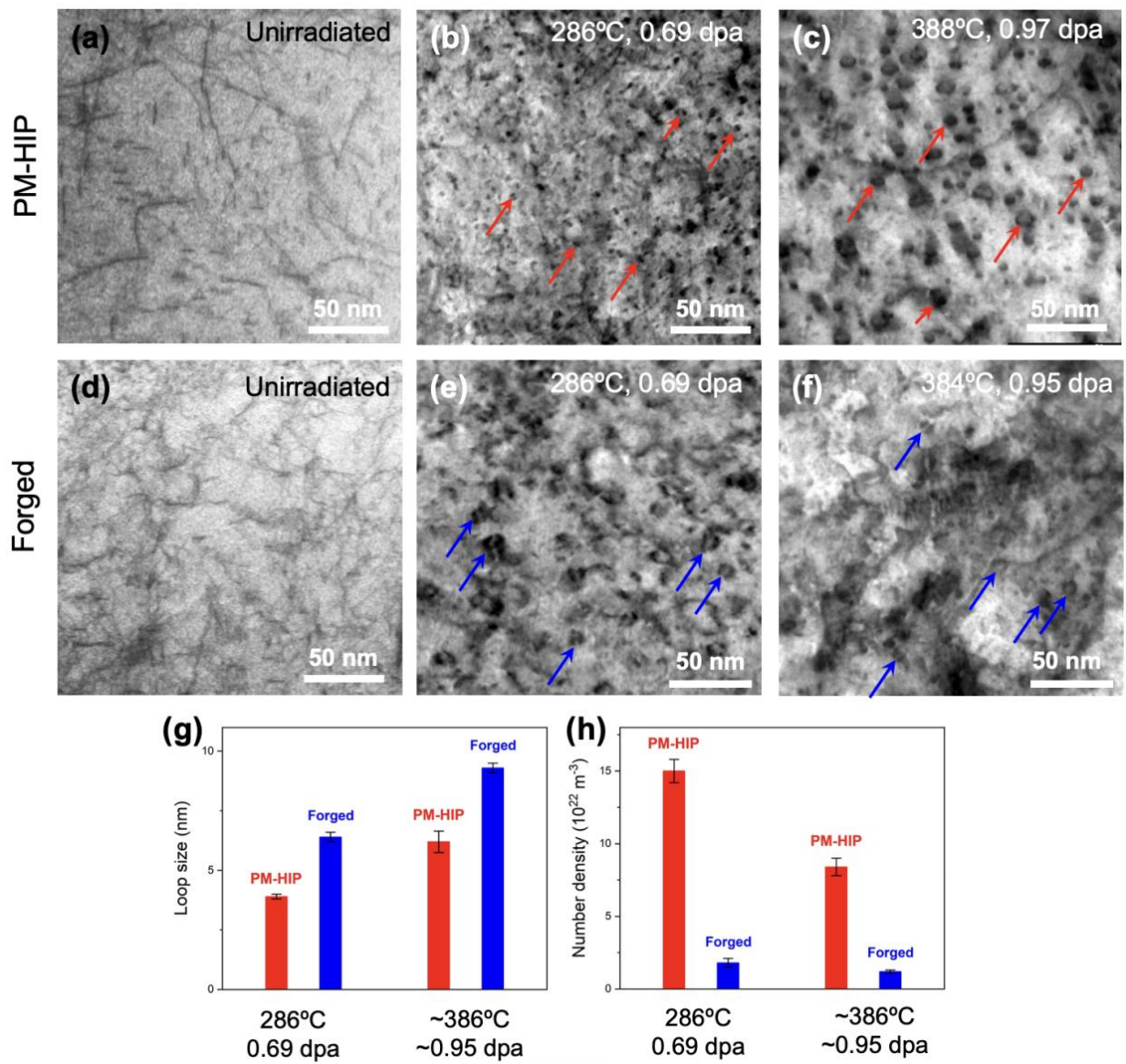

*Fig. 6 Bright field down-zone STEM images showing dislocation structures in (a) unirradiated and (b,c) neutron irradiated PM-HIP, and in (d) unirradiated and (e,f) neutron irradiated forged SA508; with quantitative comparison of dislocation loop (g) average size and (h) number density across irradiation conditions. Arrows indicate the types of features counted as dislocation loops.*

*3.4 Nanoprecipitates*

Fig. 7 shows representative APT reconstructions for Mn, Ni, and Si in unirradiated and irradiated PM-HIP and forged SA508. The reconstructions show evidence of Mn-Ni-Si-rich precipitate (MNSP) nucleation in both materials at both irradiation conditions [94]. At 0.69 dpa, 286 °C, the average nanoprecipitate sizes are statistically identical in both alloys (2.01-2.18 nm), although the PM-HIP has a higher number density than the forged (3.70 compared to 2.60 × $10^{24}$

m$^{-3}$). Consequently, the MNSPs occupy a higher volume fraction in the PM-HIP than in the forged material, ~16% compared to ~14%. The nanoprecipitates coarsen with increasing irradiation temperature; at ~0.95 dpa, ~385 ºC, the nanoprecipitates are marginally larger in diameter, but their number density and volume fraction decrease by factors of ~5 and ~3.5, respectively. The trend in which PM-HIP nanoprecipitate number density and volume fraction are greater than those of the forged material persist at the higher irradiation temperature. The nanoprecipitate sizes, number densities, and volume fractions for both alloys are quantitatively compared in Fig. 7(c-e) and Table 4.

To the authors' knowledge, Carter et al. [43] is the only other author to report on MNSP nucleation in PM-HIP RPV steels. They measure an MNSP number density of $3.73 \times 10^{24}$ m$^{-3}$ in the ferrite phase of PM-HIP SA508 neutron irradiated to ~0.1 dpa at 155 ± 10 ºC, which is nearly identical to the MNSP number density at 0.69 dpa, 286 °C in the present study. The higher dose here may have a similar effect as the lower temperature in Carter's work, ultimately leading to nearly identical MNSP number densities in both studies.

Beyond this direct comparison to PM-HIP RPV steel, there are limited reports of MNSPs in RPV steels at doses ≳0.5 dpa. Although Jiang, et al. [81] irradiate A508-3 steel to 1 dpa, they do so with charged particles (240 keV protons and 3 MeV Fe$^{13+}$ ions), and thus observed no MNSPs, since nanoprecipitation is suppressed under charged particle irradiation [65], [84], [95]–[98]. Almirall et al. [99] conduct 0.2 dpa and 1.8 dpa neutron irradiation at 300-320 ºC on various RPV steels. They report average MNSP sizes of 2.19 nm and 2.54 nm, and number densities of $0.69 \times 10^{24}$ m$^{-3}$ and $1.55 \times 10^{24}$ m$^{-3}$, for their two irradiation doses, respectively; these values are in close agreement with those reported in the present study. Almirall's findings also support the notion that nanoprecipitation saturates below ~0.5 dpa [83], which lends further support to the favorable comparison in MNSP size and number density between the present study and the other PM-HIP dataset from Carter at 0.1 dpa [43].

The APT compositions of the bulk (i.e., entire needle), matrix, and MNSPs are summarized in Table 5. Nanoprecipitates in the PM-HIP material have greater enrichment of Ni, Mn, and Si than those in the forged material at both irradiation temperatures. Accordingly, the composition of Ni, Mn, and Si remaining in the matrix is less depleted in the forged than in the PM-HIP, compared to their respective unirradiated bulk APT compositions. In both alloys, the nanoprecipitates are more enriched in solutes at the higher irradiation temperatures, although the ~0.3 dpa dose difference may also be partially responsible for the greater solute enrichment.

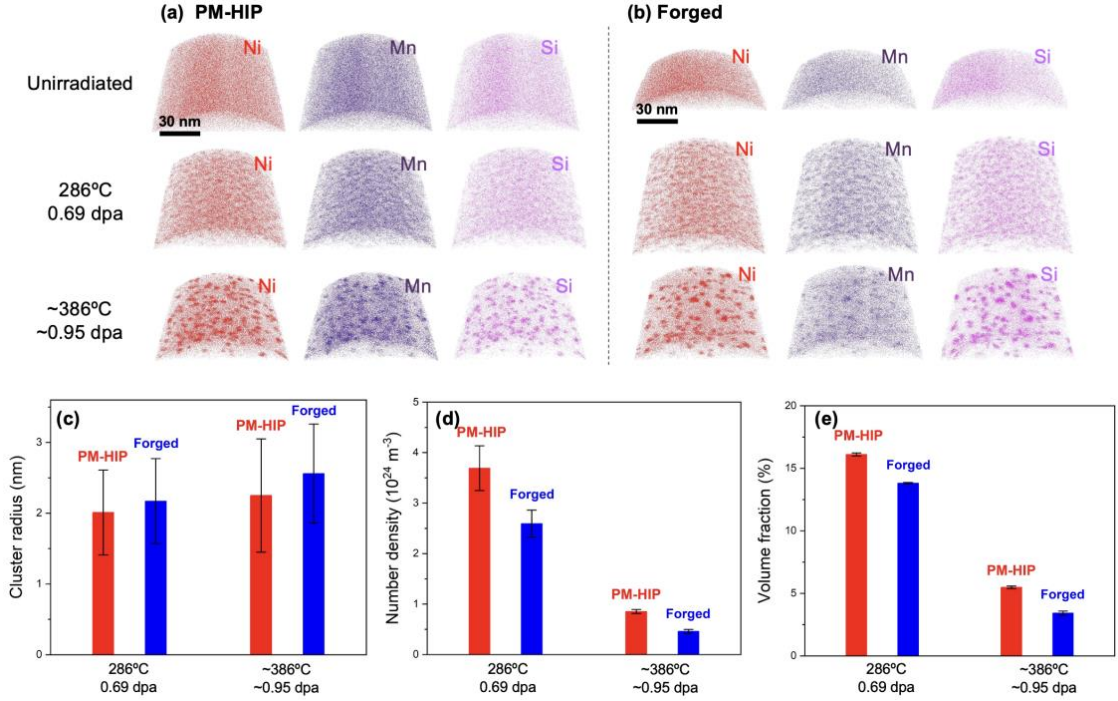

*Fig. 7 Representative APT reconstructions for Ni, Mn and Si (other species hidden for clarity of visualization) of unirradiated and irradiated (a) PM-HIP and (b) forged SA508, with quantitative analysis of MNSP (c) size, (d) number density, and (e) volume fraction.*

*Table 4 Quantitative summary of microstructural evolution, including grain and phase structure, pre-existing precipitates, dislocation loops, and MNSPs for PM-HIP and forged SA508.*

|  |  | Unirradiated |  | 286 °C, 0.69 dpa |  | ~386 °C, ~0.95 dpa |  |
|---|---|---|---|---|---|---|---|
|  |  | **PM-HIP** | **Forged** | **PM-HIP** | **Forged** | **PM-HIP** | **Forged** |
| **Gain and Phase Structure** | Bainite phase fraction | 0.74 ± 0.08 | 0.49 ± 0.07 | 0.39 ± 0.14 | 0.24 ± 0.09 | 0.30 ± 0.08 | 0.49 ± 0.11 |
|  | Ferrite phase fraction | 0.26 ± 0.08 | 0.51 ± 0.07 | 0.61 ± 0.14 | 0.76 ± 0.09 | 0.70 ± 0.08 | 0.51 ± 0.11 |
|  | Bainite grain size (μm) | 8.83 ± 1.60 | 9.78 ± 1.72 | 3.58 ± 1.36 | 4.77 ± 2.17 | 3.52 ± 0.95 | 11.0 ± 2.71 |
|  | Ferrite grain size (μm] | 3.08 ± 1.07 | 10.1 ± 1.74 | 5.56 ± 1.49 | 15.0 ± 4.21 | 8.20 ± 1.21 | 11.3 ± 2.73 |
|  | Overall grain size (μm) | 6.08 ± 0.92 | 10.1 ± 1.13 | 4.68 ± 0.82 | 10.6 ± 3.00 | 5.92 ± 0.64 | 11.2 ± 1.31 |
| **Precipitate** | Length (nm) | 49 ± 6 | 101 ± 6 | 43 ± 4 | 103 ± 7 | 48 ± 3 | 105 ± 9 |
| **Dislocation Loops** | Loop size (nm) | - | - | 3.90 ± 0.09 | 6.39 ± 0.18 | 6.19 ± 0.45 | 9.28 ± 0.18 |
|  | Number density ($10^{22}$ m$^{-3}$) | - | - | 15.04 ± 0.77 | 1.81 ± 0.29 | 8.40 ± 0.59 | 1.22 ± 0.07 |
| **MNSPs** | Radius (nm) | - | - | 2.01 ± 0.60 | 2.18 ± 0.60 | 2.25 ± 0.79 | 2.56 ± 0.70 |
|  | Number density ($10^{24}$ m$^{-3}$) | - | - | 3.70 ± 0.44 | 2.60 ± 0.26 | 0.86 ± 0.04 | 0.47 ± 0.03 |
|  | Volume fraction (%) | - | - | 16.13 ± 0.13 | 13.83 ± 0.04 | 5.52 ± 0.08 | 3.47 ± 3.47 |

***Table 5*** *Summary of APT composition measurements (wt%) of clustering species in the bulk, matrix, and in nanoprecipitates.*

| Alloy | Irradiation Condition | Bulk | | | Matrix | | | Nanoprecipitates | | |
|---|---|---|---|---|---|---|---|---|---|---|
| | | Ni | Mn | Si | Ni | Mn | Si | Ni | Mn | Si |
| **PM-HIP** | Unirradiated | 0.75 | 1.56 | 0.26 | – | – | – | – | – | – |
| | 286 °C, 0.69 dpa | 0.84 | 1.42 | 0.20 | 0.42 | 0.68 | 0.11 | 31.2 | 41.7 | 22.7 |
| | 388 °C, 0.97 dpa | 0.77 | 1.54 | 0.14 | 0.36 | 0.49 | 0.09 | 36.2 | 46.2 | 25.5 |
| **Forged** | Unirradiated | 0.39 | 1.41 | 0.37 | – | – | – | – | – | – |
| | 286 °C, 0.69 dpa | 0.44 | 1.49 | 0.32 | 0.18 | 0.75 | 0.17 | 29.3 | 38.4 | 17.9 |
| | 384 °C, 0.95 dpa | 0.51 | 1.37 | 0.41 | 0.24 | 0.52 | 0.12 | 33.7 | 42.6 | 20.4 |

## 4. Discussion

### *4.1 Structure-property relationships*

Total irradiation strengthening, $\Delta\sigma_y$, is the combination of individual strengthening contributions of each microstructural obstacle *i* to impede dislocation motion. If the features have highly differing strengths, a linear sum of the individual contributions is most accurate, but if the obstacles have similar strengths, a root-sum-square of the individual contributions is more accurate. Expressions for linear and root-sum-square are provided in Item S3 in the Supplementary Information. The strengthening contributions of individual microstructural characteristics can be determined by the Orowan dispersed barrier hardening (DBH) model [100]:

$$\Delta\sigma_{y,i} = \alpha_i M \mu b \sqrt{N_i d_i} \qquad \text{(Eq. 1)}$$

where M refers to the Taylor factor which is 3.06 for both fcc and bcc lattices [101], μ refers to the shear modulus which has a value of 79 GPa for SA508 Class 3 material [102], b refers to the Burgers vector which is 0.25 nm [103]–[105], N refers to the number densities of the obstacle, and d refers to the diameter of the obstacle. In addition, α refers to the barrier strength of the obstacle against dislocation motion and is dependent on the Poisson's ratio of the material, the dislocation core radii, obstacle size, and obstacle number density [106].

Microstructural characterization reveals dislocation loops and MNSPs are the principal contributors to hardening in the irradiated SA508 materials. For the present calculation, α = 0.33 is chosen for dislocation loops [107], [108] and α = 0.03 is chosen for MNSPs [26], [106], based on values previously used for similar obstacle distributions in similar alloys. Taking N and d values from the microstructure results reported in this study, the calculated $\Delta\sigma_y$ values are compared with experimental measurements in Fig. 8 and are also summarized in Table S1 of Item S2 in the Supplementary Information.

The linear sum approach is a close fit to experimental measurements for the PM-HIP specimens, while the root-sum-square approach significantly underestimates measured hardening. For the forged specimens, experimental measurements fall roughly halfway between the linear sum and root-sum-square calculations. This suggests that in the forged material, loops act as considerably stronger obstacles to dislocation motion than nanoprecipitates, and thus loops dominate irradiation hardening due to the low nanoprecipitates number density. Meanwhile, in the PM-HIP, loops and nanoprecipitates act more comparable in their strengthening effects on the material. So even though a single nanoprecipitate may be weaker ($\alpha = 0.03$) than a single loop ($\alpha = 0.33$), the collective effect of the high nanoprecipitate number density is a significant contribution to the overall strengthening of the alloy.

The irradiation amorphization of the oxide precipitates in the PM-HIP specimens is not included in the Orowan DBH calculation, even though irradiation amorphization can have embrittling effects [109]–[112]. If one could account for this phase transformation in the total $\Delta\sigma_y$ for PM-HIP materials, the root-sum-square calculation approach may fall into closer agreement with experimental measurements.

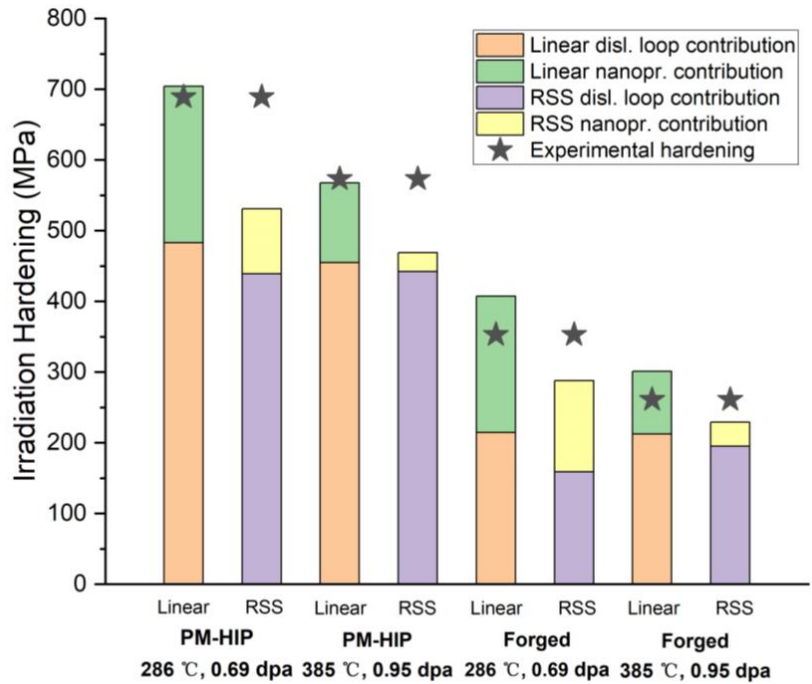

*Fig. 8* Total hardening contributions of loops and nanoprecipitates in neutron irradiated PM-HIP and forged SA508 using linear and root-sum-squared calculations, compared to experimental measurements.

*4.2 Irradiation susceptibility of PM-HIP SA508*

At low doses, microstructural features typically thought to contribute to RPV irradiation embrittlement are generally classified as either nanoprecipitates or matrix damage [113]. The former include Cu-rich precipitates (CRP), MNSPs [94], depending on bulk alloy concentration [2], [41], [43], [70], [72], [90], [99], [114]–[116]. The latter is generally defined as vacancy-type or interstitial-type point-defect clusters [90], [114]. However, at doses above ~0.1 dpa, experimental studies and atomistic simulations have shown that irradiation drives MNPs/MNSPs to agglomerate on point defect clusters [92], [117], [118]. thereby muddling the distinction between nanoprecipitates and matrix damage.

Through LWR end-of-life fluences for RPV steels, the interstitial-type clusters are typically dislocation loops <5 nm in diameter [92], [119]–[122] that remain difficult to resolve by TEM and which may form by solute-defect trapping at undersized substitutional solutes (e.g., Ni in Fe) [85]. But at higher fluences – such as the case herein – these loops can grow sufficiently large to be TEM resolvable [123], [124], though they are often decorated by Mn and Ni (and Cu, if present in the bulk material) segregation to loop boundaries, leading to heterogeneous nucleation of the Mn-Ni-rich nanoprecipitates [113], [117]. This underscores the importance of Mn and Ni in stabilizing the loop and nanoprecipitate population. The higher bulk concentration of Mn and Ni in the PM-HIP alloy (1.39 wt% and 0.79 wt%, respectively) than in the forged alloy (0.46 wt% Mn and 0.50 wt% Ni) therefore explains the higher number densities of loops and MNSPs in the PM-HIP alloy. Hence, irradiation susceptibility may be more strongly influenced by bulk alloy chemistry than fabrication or processing method.

Terentyev, et al [113] uses molecular dynamics and Metropolis Monte Carlo simulations to show that Mn enrichment to the core of dislocation loops – either alone or with other key solutes such as Ni – are principally responsible for loop strength, causing significant increases in unpinning stress. Such solute segregation to loops additionally enhances their resistance to being absorbed by gliding dislocations. It thus follows that the PM-HIP alloy having higher bulk Mn and Ni concentration than the forged material, will exhibit significant strengthening. This also supports the Orowan DBH model results showing that loops and nanoprecipitates have comparable strengthening contributions in the PM-HIP material (i.e., close agreement between linear sum approach and experimental measurements); by comparison, loops are the dominant strengthening feature in the lower-Mn, lower-Ni forged alloy.

Mechanically, the irradiated PM-HIP alloy exhibits some favorable and some unfavorable characteristics as compared to the forged alloy. Unfavorably, the PM-HIP exhibits greater irradiation-induced reduction in TE than the forged material. But at similar irradiation conditions, the two alloys have comparable UE and the PM-HIP has higher strength than the forged material. This implies that, the irradiated PM-HIP has comparable ductility as the forged material between their respective YS and UTS, which are considerably higher in the PM-HIP. In other words, under irradiation, the PM-HIP alloy exhibits superior ductility at maximum load-bearing capacity than the forged material. In addition, both the PM-HIP and forged alloys exhibit comparable approximate toughness (integrated area under stress-strain curve).

To qualify PM-HIP SA508 for irradiation-facing components in nuclear applications, the irradiation tolerance must be compared to that of conventional forgings. Microstructurally, PM-HIP SA508 appears more susceptible to nucleation of irradiation-induced dislocation loops and nanoprecipitates than its forged counterpart. This consequently leads to greater reduction in TE in the irradiated PM-HIP alloy than in the forged material. Nevertheless, the irradiated PM-HIP alloy retains some favorable mechanical behaviors, including strength and UE. Since the nature of the irradiation susceptibility is linked to bulk alloy chemistry (rather than inherent processing differences between PM-HIP and forging), future studies focusing on compositional tailoring of SA508 powders may enhance the irradiation tolerance of PM-HIP SA508. This study can guide the compositional design of SA508 powders and can also be used to motivate proper fracture toughness testing using Charpy or compact tension specimens to support a code case for qualifying PM-HIP RPV steels for nuclear applications.

## 5. Conclusions

This study provides a direct parallel comparison between PH-HIP and forged SA508 Grade 3 Class 1 low-alloy RPV steel under neutron irradiation. Both alloys are irradiated in the ATR to two conditions: 0.54-0.83 dpa at an average temperature range 266-268 ºC, and 0.95-1.00 dpa at an average temperature range 343-388 ºC. Post-irradiation characterization includes uniaxial tensile testing and microstructure characterization using SEM, S/TEM, and APT analyses. Major conclusions are as follows:

- PM-HIP SA508 undergoes greater irradiation-induced reduction in TE than forged SA508. However, UE (i.e., ductility between yield and UTS) is comparable in both PM-HIP and forged SA508. Since PM-HIP SA508 exhibits greater irradiation hardening (increase in YS and UTS), it thus has superior ductility at load-bearing capacity than the forged material, and comparable overall toughness.
- PM-HIP SA508 has a fine distribution of spherical oxide precipitates that are likely artifacts from PM-HIP processing, while forged SA508 contains needle-like precipitates. The oxides amorphize under irradiation.
- Irradiation-induced microstructural features – namely, dislocation loops and MNSPs – in PM-HIP SA508 are marginally smaller, but are present at a higher number density,

- than in the forging. Both loops and MNSPs coarsen with increasing irradiation temperature.
- The Orowan DBH model with a linear sum approach falls into close agreement with experimental measurements of $\Delta\sigma_y$ for the PM-HIP alloy, suggesting loops and MNSPs have comparable strengthening effects on the material. Whereas in the forged material, loops act as stronger obstacles than MNSPs.
- The higher bulk Mn and Ni concentration of the PM-HIP alloy – rather than inherent processing characteristics – can explain its greater susceptibility to irradiation-induced dislocation loop and MNSP nucleation, and consequently, the higher irradiation hardening. Compositional tailoring of SA508 powders may be able to further enhance the irradiation tolerance of PM-HIP SA508.

These conclusions show promise for the irradiation performance of PM-HIP SA508 relative to its forged counterpart. This work can be used to motivate fracture toughness testing as required for nuclear code qualification of PM-HIP SA508.

**CRediT Authorship Contribution Statement**

**Wen Jiang:** Investigation, Formal analysis, Writing – original draft, Writing – review & editing

**Yangyang Zhao:** Investigation, Formal analysis, Writing – original draft

**Yu Lu:** Investigation, Data curation

**Yaqiao Wu:** Investigation, Data curation, Writing – review & editing

**David Frazer:** Investigation, Data curation

**Donna P. Guillen:** Project administration, Conceptualization, Writing – review & editing

**David W. Gandy:** Project administration, Conceptualization, Funding acquisition

**Janelle P. Wharry:** Conceptualization, Funding acquisition, Writing – review & editing

**Declaration of competing interest**

The authors declare that they have no known competing financial interests or personal relationships that could have appeared to influence the work reported in this paper.

**Data availability**

Mechanical testing data is available open access in ref. [48]. Microstructural data will be made available upon request.

**Acknowledgements**

The authors thank Jeremy Burgener, Megha Dubey, and the staff at the Center for Advanced Energy Studies (CAES) for their assistance with microscopy and specimen handling; Dr. Benjamin Sutton at the Electric Power Research Institute (EPRI) for material procurement; Katie Anderson, Katelyn Baird, and Collin Knight assistance with irradiations, specimen handling, shipping, hot cell testing, and project management; Jasmyne Emerson, Sukanya Majumder, and Dr. Maria Okuniewski of Purdue University for their assistance with SEM image quantification. Funding support for this project was provided by EPRI. Irradiation experiments and post-irradiation examination were supported by the U.S. Department of Energy – Office of Nuclear Energy, through the Nuclear Science User Facilities (NSUF) contract 15-8242.

# References

[1] G. R. Odette, T. Yamamoto, T. J. Williams, R. K. Nanstad, and C. A. English, "On the history and status of reactor pressure vessel steel ductile to brittle transition temperature shift prediction models," *Journal of Nuclear Materials*, vol. 526, p. 151863, Dec. 2019, doi: 10.1016/j.jnucmat.2019.151863.

[2] G. R. Odette and R. K. Nanstad, "Predictive reactor pressure vessel steel irradiation embrittlement models: Issues and opportunities," *JOM*, vol. 61, no. 7, pp. 17–23, Jul. 2009, doi: 10.1007/s11837-009-0097-4.

[3] G. R. Odette and G. E. Lucas, "Recent progress in understanding reactor pressure vessel steel embrittlement," *Radiation Effects and Defects in Solids*, vol. 144, no. 1–4, pp. 189–231, Jun. 1998, doi: 10.1080/10420159808229676.

[4] S. Lee, S. Kim, B. Hwang, B. S. Lee, and C. G. Lee, "Effect of carbide distribution on the fracture toughness in the transition temperature region of an SA 508 steel," *Acta Mater*, vol. 50, no. 19, pp. 4755–4762, Nov. 2002, doi: 10.1016/S1359-6454(02)00313-0.

[5] K. D. Haverkamp, K. Forch, K.-H. Piehl, and W. Witte, "Effect of heat treatment and precipitation state on toughness of heavy section Mn-Mo-Ni-steel for nuclear power plants components," *Nuclear Engineering and Design*, vol. 81, no. 2, pp. 207–217, Sep. 1984, doi: 10.1016/0029-5493(84)90008-6.

[6] S. Kim, S. Lee, Y.-R. Im, H.-C. Lee, Y. J. Oh, and J. H. Hong, "Effects of alloying elements on mechanical and fracture properties of base metals and simulated heat-affected zones of SA 508 steels," *Metallurgical and Materials Transactions A*, vol. 32, no. 4, pp. 903–911, Apr. 2001, doi: 10.1007/s11661-001-0347-8.

[7] H. P. Seifert, S. Ritter, T. Shoji, Q. J. Peng, Y. Takeda, and Z. P. Lu, "Environmentally-assisted cracking behaviour in the transition region of an Alloy182/SA 508 Cl.2 dissimilar metal weld joint in simulated boiling water reactor normal water chemistry environment," *Journal of Nuclear Materials*, vol. 378, no. 2, pp. 197–210, Aug. 2008, doi: 10.1016/j.jnucmat.2008.06.034.

[8] J. Heldt and H. P. Seifert, "Stress corrosion cracking of low-alloy, reactor-pressure-vessel steels in oxygenated, high-temperature water," *Nuclear Engineering and Design*, vol. 206, no. 1, pp. 57–89, May 2001, doi: 10.1016/S0029-5493(00)00381-2.

[9] P. Soulat and P. Petrequin, "The effect of irradiation on the toughness of pressurized water reactor vessel steels under different service conditions," *Nuclear Engineering and Design*, vol. 81, no. 1, pp. 61–68, Aug. 1984, doi: 10.1016/0029-5493(84)90251-6.

[10] N. Horton and R. Sheppard, "Benefits of Hot Isostatic Pressure/Powdered Metal (HIP/PM) and Additive Manufacturing (AM) To Fabricate Advanced Energy System Components," Pittsburgh, PA, and Morgantown, WV (United States), Dec. 2016. doi: 10.2172/1417877.

[11] P. Yvon and F. Carré, "Structural materials challenges for advanced reactor systems," *Journal of Nuclear Materials*, vol. 385, no. 2, pp. 217–222, Mar. 2009, doi: 10.1016/j.jnucmat.2008.11.026.

[12]   K. L. Murty and I. Charit, "Structural materials for Gen-IV nuclear reactors: Challenges and opportunities," *Journal of Nuclear Materials*, vol. 383, no. 1–2, pp. 189–195, Dec. 2008, doi: 10.1016/j.jnucmat.2008.08.044.

[13]   P. P. Joshi, N. Kumar, and K. L. Murty, "Materials for Nuclear Reactors," in *Encyclopedia of Materials: Metals and Alloys*, Elsevier, 2022, pp. 364–376. doi: 10.1016/B978-0-12-803581-8.12070-3.

[14]   M. S. Wechsler, "Radiation embrittlement in the pressure-vessel steels of nuclear power plants," *JOM*, vol. 41, no. 12, pp. 7–14, Dec. 1989, doi: 10.1007/BF03220839.

[15]   J. Kameda, H. Takahashi, and M. Suzuki, "Residual Stress Relief and Local Embrittlement in a A533B Reactor Pressure Vessel Weldment," *International Journal of Pressure Vessels and Piping*, vol. 6, pp. 245–274, 1978.

[16]   B. S. Lee, M. C. Kim, M. W. Kim, J. H. Yoon, and J. H. Hong, "Master curve techniques to evaluate an irradiation embrittlement of nuclear reactor pressure vessels for a long-term operation," *International Journal of Pressure Vessels and Piping*, vol. 85, no. 9, pp. 593–599, Sep. 2008, doi: 10.1016/j.ijpvp.2007.08.005.

[17]   B. Z. Margolin, V. A. Nikolayev, E. V. Yurchenko, Y. A. Nikolayev, D. Y. Erak, and A. V. Nikolayeva, "Analysis of embrittlement of WWER-1000 RPV materials," *International Journal of Pressure Vessels and Piping*, vol. 89, pp. 178–186, Jan. 2012, doi: 10.1016/j.ijpvp.2011.11.003.

[18]   B. Gurovich *et al.*, "Evolution of structure and properties of VVER-1000 RPV steels under accelerated irradiation up to beyond design fluences," *Journal of Nuclear Materials*, vol. 456, pp. 23–32, 2015, doi: 10.1016/j.jnucmat.2014.09.019.

[19]   R. Chaouadi and R. Ge, "Copper precipitate hardening of irradiated RPV materials and implications on the superposition law and re-irradiation kinetics," vol. 345, pp. 65–74, 2005, doi: 10.1016/j.jnucmat.2005.05.001.

[20]   H. V. Atkinson and S. Davies, "Fundamental aspects of hot isostatic pressing: An overview," *Metallurgical and Materials Transactions A*, vol. 31, no. 12, pp. 2981–3000, Dec. 2000, doi: 10.1007/s11661-000-0078-2.

[21]   K. S. Mao, Y. Wu, C. Sun, E. Perez, and J. P. Wharry, "Laser weld-induced formation of amorphous Mn–Si precipitate in 304 stainless steel," *Materialia (Oxf)*, no. 3, pp. 174–177, Aug. 2018, doi: 10.1016/j.mtla.2018.08.012.

[22]   K. Mao, H. Wang, Y. Wu, V. Tomar, and J. P. Wharry, "Microstructure-property relationship for AISI 304/308L stainless steel laser weldment," *Materials Science and Engineering A*, vol. 721, pp. 234–243, 2018, doi: 10.1016/j.msea.2018.02.092.

[23]   K. S. Mao *et al.*, "Microstructure and microchemistry of laser welds of irradiated austenitic steels," *Mater Des*, vol. 206, p. 109764, Aug. 2021, doi: 10.1016/j.matdes.2021.109764.

[24]   D. W. Gandy, J. Shingledecker, and J. Siefert, "Overcoming Barriers for Using PM/HIP Technology to Manufacture Large Power Generation Components PM/HIP opens up a new method of

manufacturing high pressure-retaining components for use in the power-generation industry," *Advanced Materials & Processes*, vol. 170, no. 1, pp. 1–8, 2012.

[25] D. W. Gandy, C. Stover, K. Bridger, and S. Lawler, "Small Modular Reactor Vessel Manufacture/Fabrication Using PM-HIP and Electron Beam Welding Technologies," *Materials Research Proceedings (Hot Isostatic Pressing: HIP'17)*, vol. 10, pp. 224–234, 2019, doi: 10.21741/9781644900031-29.

[26] C. Clement *et al.*, "Comparing structure-property evolution for PM-HIP and forged alloy 625 irradiated with neutrons to 1 dpa," *Materials Science and Engineering: A*, vol. 857, p. 144058, Nov. 2022, doi: 10.1016/j.msea.2022.144058.

[27] C. Clement *et al.*, "Comparison of ion irradiation effects in PM-HIP and forged alloy 625," *Journal of Nuclear Materials*, vol. 558, p. 153390, Jan. 2022, doi: 10.1016/j.jnucmat.2021.153390.

[28] D. W. Gandy *et al.*, "Development of a Cobalt-free Hard-facing Alloy — NitroMaxx-PM for Nuclear Applications," in *World PM2016*, 2016.

[29] R. Ahmed *et al.*, "Single asperity nanoscratch behaviour of HIPed and cast Stellite 6 alloys," *Wear*, vol. 312, no. 1–2, pp. 70–82, 2014, doi: 10.1016/j.wear.2014.02.006.

[30] H. Yu, R. Ahmed, H. Villiers Lovelock de, and H. Davies, "Influence of Manufacturing Process and Alloying Element Content on the Tribomechanical Properties of Cobalt-Bases Alloys," *Journal of Tribology (Transactions of the ASME)*, vol. 131, no. January, pp. 011601.1-011601.12, 2009, doi: 10.1115/1.2991122.

[31] R. Ahmed, H. L. De Villiers Lovelock, S. Davies, and N. H. Faisal, "Influence of Re-HIPing on the structure-property relationships of cobalt-based alloys," *Tribol Int*, vol. 57, pp. 8–21, 2013, doi: 10.1016/j.triboint.2012.06.025.

[32] A. V. Shulga, "A comparative study of the mechanical properties and the behavior of carbon and boron in stainless steel cladding tubes fabricated by PM HIP and traditional technologies," *Journal of Nuclear Materials*, vol. 434, no. 1–3, pp. 133–140, Mar. 2013, doi: 10.1016/j.jnucmat.2012.11.008.

[33] D. W. Gandy, J. Shingledecker, and J. Siefert, "Overcoming Barriers for Using PM/HIP Technology to Manufacture Large Power Generation Components," *AM&P Technical Articles*, vol. 170, no. 1, pp. 19–23, Jan. 2012, doi: 10.31399/asm.amp.2012-01.p019.

[34] D. W. Gandy, "PM-HIP research for structural and pressuring retaining applications within the electric power industry," 2015.

[35] E. Getto *et al.*, "Thermal Aging and the Hall–Petch Relationship of PM-HIP and Wrought Alloy 625," *JOM*, vol. 71, no. 8, pp. 2837–2845, Aug. 2019, doi: 10.1007/s11837-019-03532-6.

[36] H. V. Atkinson and S. Davies, "Fundamental aspects of hot isostatic pressing: An overview," *Metallurgical and Materials Transactions A*, vol. 31, no. 12, pp. 2981–3000, Dec. 2000, doi: 10.1007/s11661-000-0078-2.

[37] H. R. Dugdale and J. B. Borradaile, "Development of hot isostatically pressed nickel based alloys for nuclear applications," *Powder Metallurgy*, vol. 56, no. 5, pp. 374–381, Dec. 2013, doi: 10.1179/1743290113Y.0000000076.

[38] D. P. Guillen, J. P. Wharry, G. Housley, C. D. Hale, J. Brookman, and D. W. Gandy, "Irradiation Experiment Design for the Evaluation of PM-HIP Alloys for Nuclear Reactors," *Nuclear Engineering and Design*, vol. 402, p. 112114, 2023, doi: https://doi.org/10.1016/j.nucengdes.2022.112114.

[39] J. Wharry *et al.*, "Materials qualification through the Nuclear Science User Facilities (NSUF): A case study on irradiated PM-HIP structural alloys," *Frontiers in Nuclear Engineering, submitted*.

[40] C. Gasparrini *et al.*, "Micromechanical testing of unirradiated and helium ion irradiated SA508 reactor pressure vessel steels: Nanoindentation vs in-situ microtensile testing," *Materials Science and Engineering: A*, vol. 796, p. 139942, Oct. 2020, doi: 10.1016/j.msea.2020.139942.

[41] K. Fukuya, "Current understanding of radiation-induced degradation in light water reactor structural materials," *J Nucl Sci Technol*, vol. 50, no. 3, pp. 213–254, Mar. 2013, doi: 10.1080/00223131.2013.772448.

[42] C. English and J. Hyde, "Radiation Damage of Reactor Pressure Vessel Steels," in *Comprehensive Nuclear Materials*, Elsevier, 2012, pp. 169–196. doi: 10.1016/B978-0-08-102865-0.00076-5.

[43] M. Carter *et al.*, "On the influence of microstructure on the neutron irradiation response of HIPed SA508 steel for nuclear applications," *Journal of Nuclear Materials*, vol. 559, p. 153435, Feb. 2022, doi: 10.1016/j.jnucmat.2021.153435.

[44] Q. Xiong *et al.*, "Characterization of microstructure of A508III/309L/308L weld and oxide films formed in deaerated high-temperature water," *Journal of Nuclear Materials*, vol. 498, pp. 227–240, Jan. 2018, doi: 10.1016/j.jnucmat.2017.10.030.

[45] S. Kim, S. Lee, Y.-R. Im, H.-C. Lee, Y. J. Oh, and J. H. Hong, "Effects of alloying elements on mechanical and fracture properties of base metals and simulated heat-affected zones of SA 508 steels," *Metallurgical and Materials Transactions A*, vol. 32, no. 4, pp. 903–911, Apr. 2001, doi: 10.1007/s11661-001-0347-8.

[46] D. P. Guillen, J. P. Wharry, G. K. Housley, C. D. Hale, J. V. Brookman, and D. W. Gandy, "Experiment design for the neutron irradiation of PM-HIP alloys for nuclear reactors," *Nuclear Engineering and Design*, vol. 402, p. 112114, Feb. 2023, doi: 10.1016/j.nucengdes.2022.112114.

[47] C. Hale, "BSU-8242 As-Run Thermal Analysis (INL/EXT-21-63578-Rev000)," Idaho Falls, ID (United States), Jul. 2021. doi: 10.2172/1813571.

[48] J. P. Wharry *et al.*, "Mechanical testing data from neutron irradiations of PM-HIP and conventionally manufactured nuclear structural alloys," *Data Brief*, vol. 48, p. 109092, Jun. 2023, doi: 10.1016/j.dib.2023.109092.

[49] H. M. Flower and T. C. Lindley, "Electron backscattering diffraction study of acicular ferrite, bainite, and martensite steel microstructures," *Materials Science and Technology*, vol. 16, no. 1, pp. 26–40, Jan. 2000, doi: 10.1179/026708300773002636.

[50] J. Mayer, L. A. Giannuzzi, T. Kamino, and J. Michael, "TEM Sample Preparation and FIB-Induced Damage," *MRS Bull*, vol. 32, no. 5, pp. 400–407, May 2007, doi: 10.1557/mrs2007.63.

[51] K. Thompson, D. Lawrence, D. J. Larson, J. D. Olson, T. F. Kelly, and B. Gorman, "In situ site-specific specimen preparation for atom probe tomography," *Ultramicroscopy*, vol. 107, no. 2–3, pp. 131–139, Feb. 2007, doi: 10.1016/j.ultramic.2006.06.008.

[52] P. J. Phillips, M. C. Brandes, M. J. Mills, and M. De Graef, "Diffraction contrast STEM of dislocations: Imaging and simulations," *Ultramicroscopy*, vol. 111, no. 9–10, pp. 1483–1487, Aug. 2011, doi: 10.1016/j.ultramic.2011.07.001.

[53] C. M. Parish, K. G. Field, A. G. Certain, and J. P. Wharry, "Application of STEM characterization for investigating radiation effects in BCC Fe-based alloys," *J Mater Res*, vol. 30, no. 9, pp. 1275–1289, May 2015, doi: 10.1557/jmr.2015.32.

[54] D. M. Norfleet, D. M. Dimiduk, S. J. Polasik, M. D. Uchic, and M. J. Mills, "Dislocation structures and their relationship to strength in deformed nickel microcrystals," *Acta Mater*, vol. 56, no. 13, pp. 2988–3001, Aug. 2008, doi: 10.1016/j.actamat.2008.02.046.

[55] P. Xiu, H. Bei, Y. Zhang, L. Wang, and K. G. Field, "STEM Characterization of Dislocation Loops in Irradiated FCC Alloys," *Journal of Nuclear Materials*, vol. 544, p. 152658, Feb. 2021, doi: 10.1016/j.jnucmat.2020.152658.

[56] M. L. Jenkins and M. A. Kirk, *Characterisation of Radiation Damage by Transmission Electron Microscopy*. CRC Press, 2000. doi: 10.1201/9781420034646.

[57] B. Yao, D. J. Edwards, and R. J. Kurtz, "TEM characterization of dislocation loops in irradiated bcc Fe-based steels," *Journal of Nuclear Materials*, vol. 434, no. 1–3, pp. 402–410, Mar. 2013, doi: 10.1016/j.jnucmat.2012.12.002.

[58] J. E. Nathaniel *et al.*, "Toward high-throughput defect density quantification: A comparison of techniques for irradiated samples," *Ultramicroscopy*, vol. 206, p. 112820, Nov. 2019, doi: 10.1016/j.ultramic.2019.112820.

[59] P. Xiu *et al.*, "Dislocation loop evolution and radiation hardening in nickel-based concentrated solid solution alloys," *Journal of Nuclear Materials*, vol. 538, p. 152247, Sep. 2020, doi: 10.1016/j.jnucmat.2020.152247.

[60] K. G. Field, S. A. Briggs, K. Sridharan, Y. Yamamoto, and R. H. Howard, "Dislocation loop formation in model FeCrAl alloys after neutron irradiation below 1 dpa," *Journal of Nuclear Materials*, vol. 495, pp. 20–26, Nov. 2017, doi: 10.1016/j.jnucmat.2017.07.061.

[61] R. F. Egerton, *Electron Energy-Loss Spectroscopy in the Electron Microscope*. Boston, MA: Springer US, 2011. doi: 10.1007/978-1-4419-9583-4.

[62] D. Vaumousse, A. Cerezo, and P. J. Warren, "A procedure for quantification of precipitate microstructures from three-dimensional atom probe data," *Ultramicroscopy*, vol. 95, pp. 215–221, May 2003, doi: 10.1016/S0304-3991(02)00319-4.

[63] M. Bachhav, G. Robert Odette, and E. A. Marquis, "α' precipitation in neutron-irradiated Fe–Cr alloys," *Scr Mater*, vol. 74, pp. 48–51, Mar. 2014, doi: 10.1016/j.scriptamat.2013.10.001.

[64] M. J. Swenson and J. P. Wharry, "Collected data set size considerations for atom probe cluster analysis," *Microscopy and Microanalysis*, vol. 22, no. S3, pp. 690–691, 2016, doi: 10.1017/S143192761600430X.

[65] M. J. Swenson and J. P. Wharry, "The comparison of microstructure and nanocluster evolution in proton and neutron irradiated Fe-9%Cr ODS steel to 3 dpa at 500C," *Journal of Nuclear Materials*, vol. 467, pp. 97–112, 2015, doi: 10.1016/j.jnucmat.2015.09.022.

[66] M. Lambrecht, L. Malerba, and A. Almazouzi, "Influence of different chemical elements on irradiation-induced hardening embrittlement of RPV steels," *Journal of Nuclear Materials*, vol. 378, no. 3, pp. 282–290, Sep. 2008, doi: 10.1016/j.jnucmat.2008.06.030.

[67] T. S. Byun and K. Farrell, "Irradiation hardening behavior of polycrystalline metals after low temperature irradiation," *Journal of Nuclear Materials*, vol. 326, no. 2–3, pp. 86–96, Mar. 2004, doi: 10.1016/j.jnucmat.2003.12.012.

[68] Y. Liu *et al.*, "Machine learning predictions of irradiation embrittlement in reactor pressure vessel steels," *NPJ Comput Mater*, vol. 8, no. 1, p. 85, Apr. 2022, doi: 10.1038/s41524-022-00760-4.

[69] Y. Liu, D. Morgan, T. Yamamoto, and G. R. Odette, "Characterizing the flux effect on the irradiation embrittlement of reactor pressure vessel steels using machine learning," *Acta Mater*, vol. 256, p. 119144, Sep. 2023, doi: 10.1016/j.actamat.2023.119144.

[70] G. R. Odette *, T. Yamamoto, and D. Klingensmith, "On the effect of dose rate on irradiation hardening of RPV steels," *Philosophical Magazine*, vol. 85, no. 4–7, pp. 779–797, Feb. 2005, doi: 10.1080/14786430412331319910.

[71] H.-H. Jin, C. Shin, D. H. Kim, K. H. Oh, and J. Kwon, "Evolution of a Needle Shaped Carbide in SA508 Gr3 Steel," *ISIJ International*, vol. 48, no. 12, pp. 1810–1812, 2008, doi: 10.2355/isijinternational.48.1810.

[72] E. A. Marquis, "Atom probe tomography applied to the analysis of irradiated microstructures," *J Mater Res*, vol. 30, no. 9, pp. 1222–1230, May 2015, doi: 10.1557/jmr.2014.398.

[73] B. Duan *et al.*, "The effect of the initial microstructure in terms of sink strength on the ion-irradiation-induced hardening of ODS alloys studied by nanoindentation," *Journal of Nuclear Materials*, vol. 495, pp. 118–127, Nov. 2017, doi: 10.1016/j.jnucmat.2017.08.014.

[74] A. F. Rowcliffe, L. K. Mansur, D. T. Hoelzer, and R. K. Nanstad, "Perspectives on radiation effects in nickel-base alloys for applications in advanced reactors," *Journal of Nuclear Materials*, vol. 392, no. 2, pp. 341–352, Jul. 2009, doi: 10.1016/j.jnucmat.2009.03.023.

[75] M. H. Bocanegra-Bernal, "Hot Isostatic Pressing (HIP) technology and its applications to metals and ceramics," *J Mater Sci*, vol. 39, no. 21, pp. 6399–6420, Nov. 2004, doi: 10.1023/B:JMSC.0000044878.11441.90.

[76]   T. Fujita, J. Hirabayashi, Y. Katayama, F. Kano, and H. Watanabe, "Contribution of dislocation loop to radiation-hardening of RPV steels studied by STEM/EDS with surveillance test pieces," *Journal of Nuclear Materials*, vol. 572, p. 154055, Dec. 2022, doi: 10.1016/j.jnucmat.2022.154055.

[77]   F. Bergner *et al.*, "Contributions of Cu-rich clusters, dislocation loops and nanovoids to the irradiation-induced hardening of Cu-bearing low-Ni reactor pressure vessel steels," *Journal of Nuclear Materials*, vol. 461, pp. 37–44, Jun. 2015, doi: 10.1016/j.jnucmat.2015.02.031.

[78]   E. A. Kuleshova *et al.*, "Mechanisms of radiation embrittlement of VVER-1000 RPV steel at irradiation temperatures of (50–400)°C," *Journal of Nuclear Materials*, vol. 490, pp. 247–259, Jul. 2017, doi: 10.1016/j.jnucmat.2017.04.035.

[79]   E. A. Kuleshova *et al.*, "Specific Features of Structural-Phase State and Properties of Reactor Pressure Vessel Steel at Elevated Irradiation Temperature," *Science and Technology of Nuclear Installations*, vol. 2017, pp. 1–12, 2017, doi: 10.1155/2017/1064182.

[80]   M. Hernández-Mayoral and D. Gómez-Briceño, "Transmission electron microscopy study on neutron irradiated pure iron and RPV model alloys," *Journal of Nuclear Materials*, vol. 399, no. 2–3, pp. 146–153, Apr. 2010, doi: 10.1016/j.jnucmat.2009.11.013.

[81]   J. Jiang *et al.*, "Microstructural evolution of RPV steels under proton and ion irradiation studied by positron annihilation spectroscopy," *Journal of Nuclear Materials*, vol. 458, pp. 326–334, Mar. 2015, doi: 10.1016/j.jnucmat.2014.12.113.

[82]   X. Ma, Q. Zhang, L. Song, W. Zhang, M. She, and F. Zhu, "Microstructure Evolution of Reactor Pressure Vessel A508-3 Steel under High-Dose Heavy Ion Irradiation," *Crystals (Basel)*, vol. 12, no. 8, p. 1091, Aug. 2022, doi: 10.3390/cryst12081091.

[83]   B. Gómez-Ferrer, C. Heintze, and C. Pareige, "On the role of Ni, Si and P on the nanostructural evolution of FeCr alloys under irradiation," *Journal of Nuclear Materials*, vol. 517, pp. 35–44, Apr. 2019, doi: 10.1016/j.jnucmat.2019.01.040.

[84]   C. Pareige, V. Kuksenko, and P. Pareige, "Behaviour of P, Si, Ni impurities and Cr in self ion irradiated Fe–Cr alloys – Comparison to neutron irradiation," *Journal of Nuclear Materials*, vol. 456, pp. 471–476, Jan. 2015, doi: 10.1016/j.jnucmat.2014.10.024.

[85]   P. H. Warren *et al.*, "The role of Cr, P, and N solutes on the irradiated microstructure of bcc Fe," *Journal of Nuclear Materials*, vol. 583, p. 154531, Sep. 2023, doi: 10.1016/j.jnucmat.2023.154531.

[86]   S. Shen, F. Chen, X. Tang, G. Ge, J. Gao, and Z. Sun, "Irradiation damage and swelling of carbon-doped Fe38Mn40Ni11Al4Cr7 high-entropy alloys under heavy ion irradiation at elevated temperature," *J Mater Sci*, vol. 55, no. 36, pp. 17218–17231, Dec. 2020, doi: 10.1007/s10853-020-05229-7.

[87]   W.-Y. Chen, X. Liu, Y. Chen, J.-W. Yeh, K.-K. Tseng, and K. Natesan, "Irradiation effects in high entropy alloys and 316H stainless steel at 300 °C," *Journal of Nuclear Materials*, vol. 510, pp. 421–430, Nov. 2018, doi: 10.1016/j.jnucmat.2018.08.031.

[88] N. A. P. K. Kumar, C. Li, K. J. Leonard, H. Bei, and S. J. Zinkle, "Microstructural stability and mechanical behavior of FeNiMnCr high entropy alloy under ion irradiation," *Acta Mater*, vol. 113, pp. 230–244, Jul. 2016, doi: 10.1016/j.actamat.2016.05.007.

[89] Z. Su *et al.*, "The effect of interstitial carbon atoms on defect evolution in high entropy alloys under helium irradiation," *Acta Mater*, vol. 233, p. 117955, Jul. 2022, doi: 10.1016/j.actamat.2022.117955.

[90] G. R. Odette, "On the dominant mechanism of irradiation embrittlement of reactor pressure vessel steels," *Scripta Metallurgica*, vol. 17, no. 10, pp. 1183–1188, Oct. 1983, doi: 10.1016/0036-9748(83)90280-6.

[91] Y. Nagai *et al.*, "Positron annihilation study of vacancy-solute complex evolution in Fe-based alloys," *Phys Rev B*, vol. 67, no. 22, p. 224202, Jun. 2003, doi: 10.1103/PhysRevB.67.224202.

[92] E. Meslin *et al.*, "Characterization of neutron-irradiated ferritic model alloys and a RPV steel from combined APT, SANS, TEM and PAS analyses," *Journal of Nuclear Materials*, vol. 406, no. 1, pp. 73–83, Nov. 2010, doi: 10.1016/j.jnucmat.2009.12.021.

[93] S. C. Glade, B. D. Wirth, G. R. Odette, and P. Asoka-Kumar, "Positron annihilation spectroscopy and small angle neutron scattering characterization of nanostructural features in high-nickel model reactor pressure vessel steels," *Journal of Nuclear Materials*, vol. 351, no. 1–3, pp. 197–208, Jun. 2006, doi: 10.1016/j.jnucmat.2006.02.012.

[94] J.-H. Ke and B. W. Spencer, "Cluster dynamics modeling of Mn-Ni-Si precipitates coupled with radiation-induced segregation in low-Cu reactor pressure vessel steels," *Journal of Nuclear Materials*, vol. 569, p. 153910, Oct. 2022, doi: 10.1016/j.jnucmat.2022.153910.

[95] Y. Pachaury, T. Kumagai, J. P. Wharry, and A. El-Azab, "A data science approach for analysis and reconstruction of spinodal-like composition fields in irradiated FeCrAl alloys," *Acta Mater*, vol. 234, p. 118019, Aug. 2022, doi: 10.1016/j.actamat.2022.118019.

[96] C. Pareige *et al.*, "Solute rich cluster formation and Cr precipitation in irradiated Fe-Cr-(Ni,Si,P) alloys: Ion and neutron irradiation," *Journal of Nuclear Materials*, vol. 572, p. 154060, Dec. 2022, doi: 10.1016/j.jnucmat.2022.154060.

[97] M. J. Swenson and J. P. Wharry, "Rate Theory Model of Irradiation-Induced Solute Clustering in b.c.c. Fe-Based Alloys," *JOM*, vol. 72, no. 11, pp. 4017–4027, Nov. 2020, doi: 10.1007/s11837-020-04365-4.

[98] M. J. Swenson and J. P. Wharry, "Nanocluster irradiation evolution in Fe-9%Cr ODS and ferritic-martensitic alloys," *Journal of Nuclear Materials*, vol. 496, pp. 24–40, 2017, doi: 10.1016/j.jnucmat.2017.08.045.

[99] N. Almirall, P. B. Wells, T. Yamamoto, K. Yabuuchi, A. Kimura, and G. R. Odette, "On the use of charged particles to characterize precipitation in irradiated reactor pressure vessel steels with a wide range of compositions," *Journal of Nuclear Materials*, vol. 536, p. 152173, Aug. 2020, doi: 10.1016/j.jnucmat.2020.152173.

[100] G. S. Was, Ed., "Irradiation Hardening and Deformation," in *Fundamentals of Radiation Materials Science: Metals and Alloys*, Berlin, Heidelberg: Springer Berlin Heidelberg, 2007, pp. 581–642. doi: 10.1007/978-3-540-49472-0_12.

[101] R. E. Stoller and S. J. Zinkle, "On the relationship between uniaxial yield strength and resolved shear stress in polycrystalline materials," *Journal of Nuclear Materials*, vol. 283–287, pp. 349–352, Dec. 2000, doi: 10.1016/S0022-3115(00)00378-0.

[102] Y.-M. Cheong, J.-H. Kim, J.-H. Hong, and H.-K. Jung, "Dynamic elastic constants of Weld HAZ of SA 508 Class 3 steel using resonant ultrasound spectroscopy," *IEEE Transactions on Ultrasonic Ferroelectric Frequency Control*, vol. 47, no. 3, pp. 559–564, 2000.

[103] H. Liu and Q. Li, "Dislocation density evaluation of three commercial SA508Gr.3 steels for reactor pressure vessel," *IOP Conf Ser Mater Sci Eng*, vol. 490, p. 022019, Apr. 2019, doi: 10.1088/1757-899X/490/2/022019.

[104] S. Kotrechko, V. Dubinko, N. Stetsenko, D. Terentyev, X. He, and M. Sorokin, "Temperature dependence of irradiation hardening due to dislocation loops and precipitates in RPV steels and model alloys," *Journal of Nuclear Materials*, vol. 464, pp. 6–15, Sep. 2015, doi: 10.1016/j.jnucmat.2015.04.014.

[105] K. Maruyama, K. Sawada, and J. Koike, "Strengthening Mechanisms of Creep Resistant Tempered Martensitic Steel," *ISIJ International*, vol. 41(6), p. 641, Jan. 2001.

[106] M. J. Swenson, C. K. Dolph, and J. P. Wharry, "The effects of oxide evolution on mechanical properties in proton- and neutron-irradiated Fe-9%Cr ODS steel," *Journal of Nuclear Materials*, vol. 479, pp. 426–435, Oct. 2016, doi: 10.1016/j.jnucmat.2016.07.022.

[107] K. S. Mao *et al.*, "Grain orientation dependence of nanoindentation and deformation-induced martensitic phase transformation in neutron irradiated AISI 304L stainless steel," *Materialia (Oxf)*, vol. 5, p. 100208, Mar. 2019, doi: 10.1016/j.mtla.2019.100208.

[108] G. E. Lucas, "The evolution of mechanical property change in irradiated austenitic stainless steels," *Journal of Nuclear Materials*, vol. 206, no. 2–3, pp. 287–305, Nov. 1993, doi: 10.1016/0022-3115(93)90129-M.

[109] E. C. Kim, H. D. Nam, D. G. Park, J. H. Hong, and J. H. Lee, "Magnetic Properties of Neutron-Irradiated RPV Steel by Mössbauer Spectroscopy," *Hyperfine Interact*, vol. 139/140, no. 1/4, pp. 479–483, 2002, doi: 10.1023/A:1021254207855.

[110] A. Bhattacharya *et al.*, "Radiation induced amorphization of carbides in additively manufactured and conventional ferritic-martensitic steels: In-situ experiments on extraction replicas," *Journal of Nuclear Materials*, vol. 563, p. 153646, May 2022, doi: 10.1016/j.jnucmat.2022.153646.

[111] N. Baluc, R. Schäublin, C. Bailat, F. Paschoud, and M. Victoria, "The mechanical properties and microstructure of the OPTIMAX series of low activation ferritic–martensitic steels," *Journal of Nuclear Materials*, vol. 283–287, pp. 731–735, Dec. 2000, doi: 10.1016/S0022-3115(00)00282-8.

[112] L. Zhou *et al.*, "Research Progress of Steels for Nuclear Reactor Pressure Vessels," *Materials*, vol. 15, no. 24, p. 8761, Dec. 2022, doi: 10.3390/ma15248761.

[113] D. Terentyev, X. He, G. Bonny, A. Bakaev, E. Zhurkin, and L. Malerba, "Hardening due to dislocation loop damage in RPV model alloys: Role of Mn segregation," *Journal of Nuclear Materials*, vol. 457, pp. 173–181, Feb. 2015, doi: 10.1016/j.jnucmat.2014.11.023.

[114] G. R. Odette and G. E. Lucas, "Embrittlement of nuclear reactor pressure vessels," *JOM*, vol. 53, no. 7, pp. 18–22, Jul. 2001, doi: 10.1007/s11837-001-0081-0.

[115] T. Toyama, T. Yamamoto, N. Ebisawa, K. Inoue, Y. Nagai, and G. R. Odette, "Effects of neutron flux on irradiation-induced hardening and defects in RPV steels studied by positron annihilation spectroscopy," *Journal of Nuclear Materials*, vol. 532, p. 152041, Apr. 2020, doi: 10.1016/j.jnucmat.2020.152041.

[116] M. K. Miller, K. A. Powers, R. K. Nanstad, and P. Efsing, "Atom probe tomography characterizations of high nickel, low copper surveillance RPV welds irradiated to high fluences," *Journal of Nuclear Materials*, vol. 437, no. 1–3, pp. 107–115, Jun. 2013, doi: 10.1016/j.jnucmat.2013.01.312.

[117] G. Bonny *et al.*, "On the thermal stability of late blooming phases in reactor pressure vessel steels: An atomistic study," *Journal of Nuclear Materials*, vol. 442, no. 1–3, pp. 282–291, Nov. 2013, doi: 10.1016/j.jnucmat.2013.08.018.

[118] G. Bonny, D. Terentyev, E. E. Zhurkin, and L. Malerba, "Monte Carlo study of decorated dislocation loops in FeNiMnCu model alloys," *Journal of Nuclear Materials*, vol. 452, no. 1–3, pp. 486–492, Sep. 2014, doi: 10.1016/j.jnucmat.2014.05.051.

[119] J. Kočík, E. Keilová, J. Čížek, and I. Procházka, "TEM and PAS study of neutron irradiated VVER-type RPV steels," *Journal of Nuclear Materials*, vol. 303, no. 1, pp. 52–64, May 2002, doi: 10.1016/S0022-3115(02)00800-0.

[120] G. Maussner, L. Scharf, R. Langer, and B. Gurovich, "Microstructure alterations in the base material, heat affected zone and weld metal of a 440-VVER-reactor pressure vessel caused by high fluence irradiation during long term operation; material: 15 Ch2MFA ≈0.15 C–2.5 Cr–0.7 Mo–0.3 V," *Nuclear Engineering and Design*, vol. 193, no. 3, pp. 359–376, Oct. 1999, doi: 10.1016/S0029-5493(99)00192-2.

[121] E. A. Kuleshova, B. A. Gurovich, Ya. I. Shtrombakh, D. Yu. Erak, and O. V. Lavrenchuk, "Comparison of microstructural features of radiation embrittlement of VVER-440 and VVER-1000 reactor pressure vessel steels," *Journal of Nuclear Materials*, vol. 300, no. 2–3, pp. 127–140, Feb. 2002, doi: 10.1016/S0022-3115(01)00752-8.

[122] B. A. Gurovich, E. A. Kuleshova, Ya. I. Shtrombakh, D. Yu. Erak, A. A. Chernobaeva, and O. O. Zabusov, "Fine structure behaviour of VVER-1000 RPV materials under irradiation," *Journal of Nuclear Materials*, vol. 389, no. 3, pp. 490–496, Jun. 2009, doi: 10.1016/j.jnucmat.2009.02.002.

[123] K. Fujii, K. Fukuya, N. Nakata, K. Hono, Y. Nagai, and M. Hasegawa, "Hardening and microstructural evolution in A533B steels under high-dose electron irradiation," *Journal of Nuclear Materials*, vol. 340, no. 2–3, pp. 247–258, Apr. 2005, doi: 10.1016/j.jnucmat.2004.12.008.

[124] T. Hamaoka, Y. Satoh, and H. Matsui, "One-dimensional motion of self-interstitial atom clusters in A533B steel observed using a high-voltage electron microscope," *Journal of Nuclear Materials*, vol. 399, no. 1, pp. 26–31, Apr. 2010, doi: 10.1016/j.jnucmat.2009.12.014.

# SUPPLEMENTARY INFORMATION FOR:

# Comparison of PM-HIP to Forged SA508 Pressure Vessel Steel Under High-Dose Neutron Irradiation

Wen Jiang [a+], Yangyang Zhao [a+], Yu Lu [b,c], Yaqiao Wu [b,c], David Frazer [d], Donna P. Guillen [d], David W. Gandy [e], Janelle P. Wharry [a]

[a] *School of Materials Engineering, Purdue University, West Lafayette, IN, USA*

[b] *Micron School of Materials Science & Engineering, Boise State University, Boise, ID, USA*

[c] *Center for Advanced Energy Studies, Idaho Falls, ID, USA*

[d] *Idaho National Laboratory, Idaho Falls, ID, USA*

[e] *Electric Power Research Institute, Charlotte, NC, USA*

+ These authors have equivalent contributions as first author

**Item S1: Calculation of Cluster Size & Number Density from APT**

The cluster size ($r$) was defined as the spherical volume equivalent radius given by:

$$r = \left(\frac{3}{4\pi} \frac{N_{\text{atoms}}}{\rho_{\text{th}}\eta}\right)^{\frac{1}{3}} \quad \text{(Eq. 1)}$$

where $N_{\text{atoms}}$ is the number of detected atoms within the cluster, $\rho_{\text{th}}$ is the theoretical atomic density of the cluster (taken to be the same as that of the ferrite matrix, 84.3 atom nm$^{-3}$), and $\eta$ is the detection efficiency of 0.36 of the LEAP instrument used. The volume fraction ($\phi$) of the clusters was determined from:

$$\phi = \frac{\sum N_{\text{atoms}}}{N_{\text{total}}} \quad \text{(Eq. 2)}$$

where $\sum N_{\text{atoms}}$ is the sum total number of detected atoms within all the clusters in the reconstructed volume and $N_{\text{total}}$ is the total number of detected atoms in the reconstructed volume. The number density ($n_v$) of the clusters was estimated from:

$$n_v = \frac{N \rho_{\text{th}} \eta}{N_{\text{total}}} \quad \text{(Eq. 3)}$$

where $N$ is the number of identified clusters in the reconstructed volume.

**Item S2: Additional Figures & Tables**

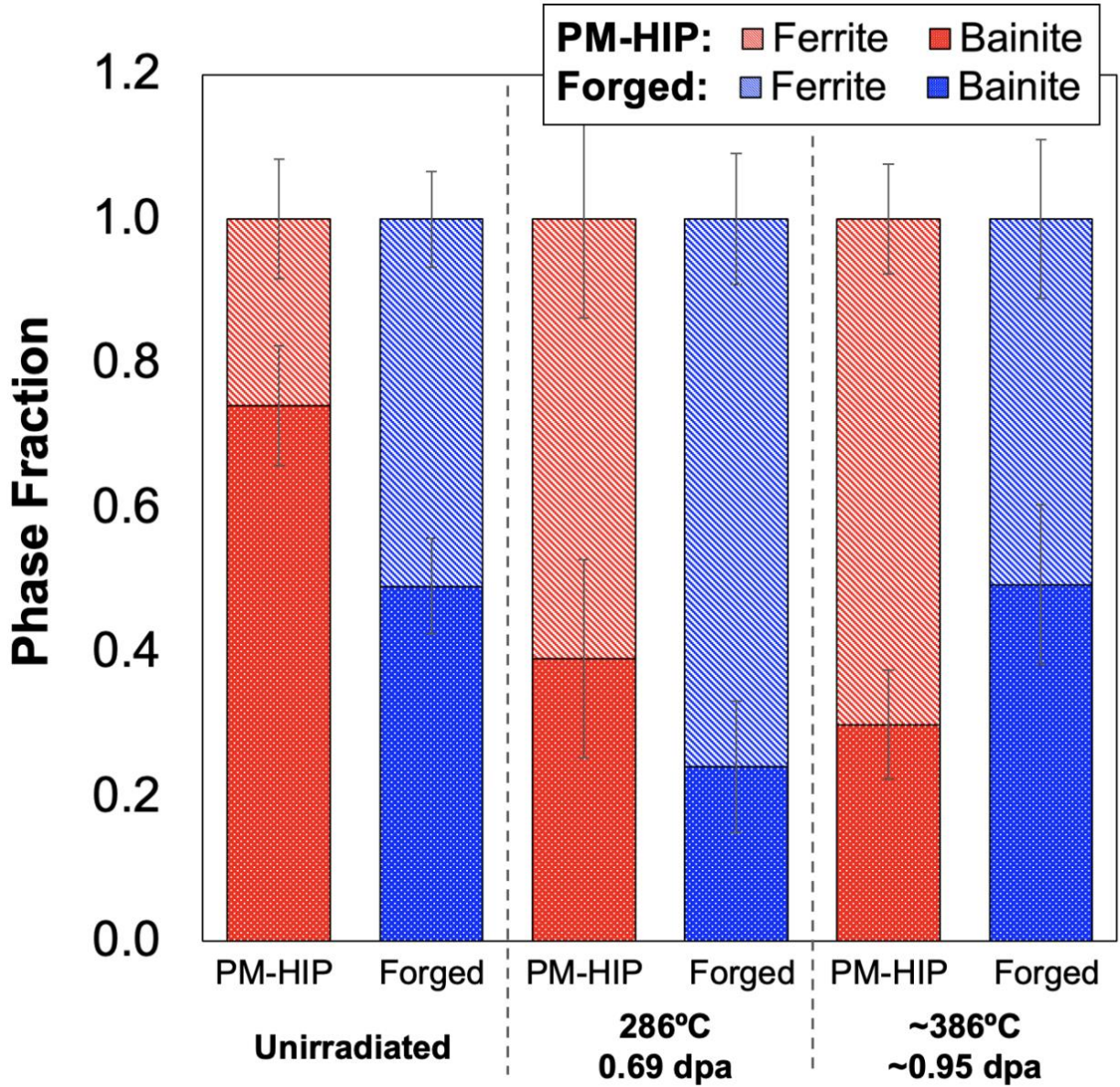

***Fig. S1*** *Evolution of ferrite and bainite phase fraction in PM-HIP and forged SA508 throughout all irradiation conditions studied.*

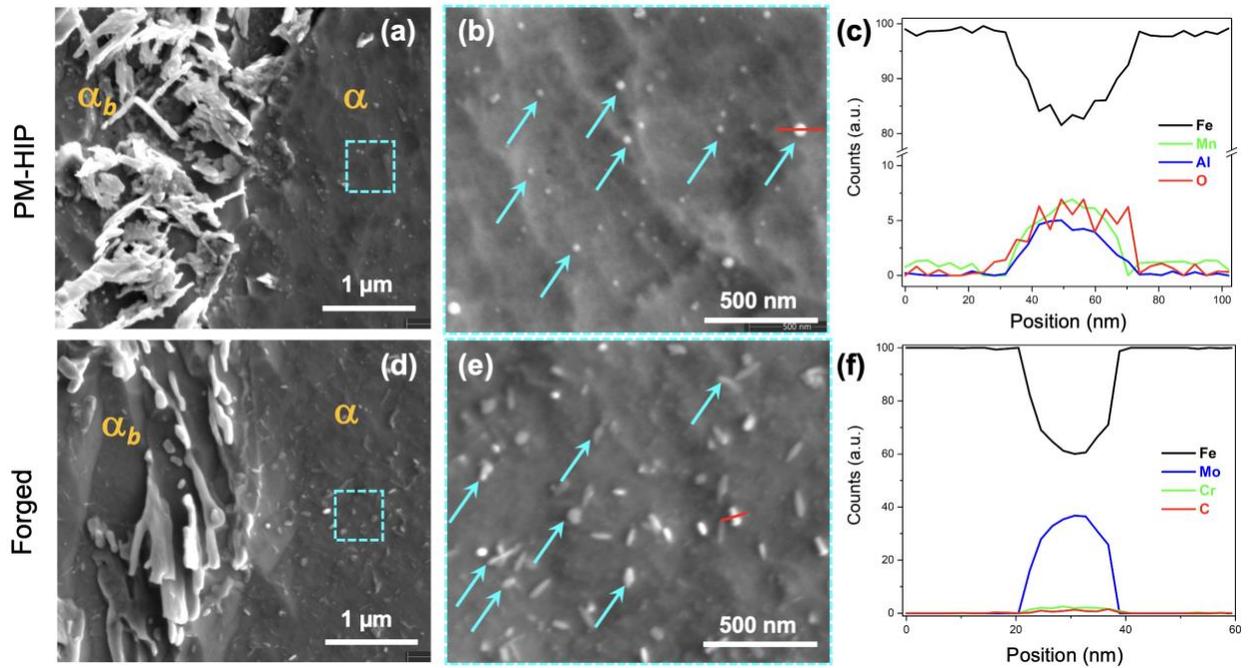

***Fig. S2*** *Scanning electron microscope (SEM) image of a ferrite-bainite phase boundary in unirradiated (a) PM-HIP and (d) forged SA508. Boxed region is shown at higher magnification revealing (b) spherical precipitates in PM-HIP and (e) needle-like precipitates in forged. SEM energy dispersive x-ray spectroscopy (EDS) line scans over selected precipitates [red line marked in (b,e)] reveal the composition of the precipitates is (c) Mn-Al oxide in PM-HIP and (f) Mo-Cr carbide in forged.*

*Table S1 Orowan dispersed barrier hardening calculations in neutron irradiated PM-HIP and forged SA508. Values in parentheses represent percent differences between the calculated and measured strengthening.*

| Irradiation hardening, $\Delta\sigma_y$ (MPa) | | 286 °C, 0.69 dpa | | ~386 °C, ~0.95 dpa | |
|---|---|---|---|---|---|
| | | **PM-HIP** | **Forged** | **PM-HIP** | **Forged** |
| $\Delta\sigma_{y,linear}$ | Dislocation loop contribution | 483 | 214 | 455 | 212 |
| | Nanoprecipitate contribution | 221 | 193 | 113 | 89 |
| | Total | 704 (2.2) | 408 (16) | 568 (-0.9) | 301 (15) |
| $\Delta\sigma_{y,rss}$ | Dislocation loop contribution | 439 | 159 | 442 | 195 |
| | Nanoprecipitate contribution | 92 | 129 | 27 | 34 |
| | Total | 531 (-23) | 288 (-18) | 469 (-18) | 230 (-12) |
| Experimentally measured (tensile) | | 689 | 353 | 573 | 261 |

## Item S3:  Strengthening Contribution Equations

Total irradiation strengthening, $\Delta\sigma_y$, is the combination of individual strengthening contributions of all $i$ microstructural obstacles to dislocation motion.

If the features have highly differing strengths, a linear sum of the individual contributions is most accurate:

$$\Delta\sigma_{y,linear} = \sum_i \Delta\sigma_{y,i} \qquad (\text{Eq. 4})$$

where $\Delta\sigma_{y,i}$ is the strengthening contribution of obstacle type $i$.

But if the obstacles have similar strengths, a root-sum-square of the individual contributions is more accurate:

$$\Delta\sigma_{y,rss} = \sqrt{\sum_i (\Delta\sigma_{y,i})^2} \qquad (\text{Eq. 5})$$